\documentclass{emulateapj}

\newcommand\eq{\begin{equation}}
\newcommand\eeq{\end{equation}}
\newcommand\eqn{\begin{eqnarray}}
\newcommand\eeqn{\end{eqnarray}}
\newbox\grsign \setbox\grsign=\hbox{$>$} \newdimen\grdimen
\grdimen=\ht\grsign
\newbox\simlessbox \newbox\simgreatbox
\setbox\simgreatbox=\hbox{\raise.5ex\hbox{$>$}\llap
     {\lower.5ex\hbox{$\sim$}}}\ht1=\grdimen\dp1=0pt
\setbox\simlessbox=\hbox{\raise.5ex\hbox{$<$}\llap
     {\lower.5ex\hbox{$\sim$}}}\ht2=\grdimen\dp2=0pt

\newbox\simppropto
\setbox\simppropto=\hbox{\raise.5ex\hbox{$\sim$}\llap
     {\lower.5ex\hbox{$\propto$}}}\ht2=\grdimen\dp2=0pt

\newcommand\kms{km s$^{-1}$}

\newcommand\sloanr{$r^\prime$}
\newcommand\sloang{$g^\prime$}

\begin{document}

\shorttitle{Witnessing the formation of a galaxy cluster}
\shortauthors{Carrasco et al.}

\title{Witnessing the formation of a galaxy cluster at $z=0.485$: optical
and  X-ray properties of RX J1117.4$+$0743
([VMF~98]~097)\altaffilmark{1}}
\altaffiltext{1}{Based on observations obtained at the Gemini
Observatory,  which is operated by the Association of Universities for
Research in  Astronomy, Inc., under a cooperative agreement with the NSF
on behalf of the Gemini partnership: the National Science Foundation
(United States), the Particle Physics and Astronomy Research Council
(United Kingdom), the National Research Council (Canada), CONICYT
(Chile), the Australian Research Council (Australia), CNPq (Brazil) and
CONICET (Argentina); Gemini program ID is GS-2003A-SV-206}
\author{E. R. Carrasco\altaffilmark{2}, E. S. Cypriano\altaffilmark{3},
G. B. Lima Neto\altaffilmark{4},
H. Cuevas\altaffilmark{5}, L. Sodr\'e Jr.\altaffilmark{4},
C. Mendes de Oliveira\altaffilmark{4},
and A. Ramirez\altaffilmark{5}}
\altaffiltext{2}{Gemini Observatory/AURA, Southern Operations Center,
Casilla 603,  La Serena, Chile}
\altaffiltext{3}{Department of Physics and Astronomy, University College
London,  London WC1E 6BT, UK}
\altaffiltext{4}{Departamento de Astronomia, Instituto de Astronomia,
Geof\'{\i}sica e Ci\^encias  Atmosf\'ericas, Universidade de S\~ao Paulo,
Rua do Mat\~ao 1226,  Cidade Universit\'aria, 05508-090, S\~ao Paulo,
Brazil}
\altaffiltext{5}{Departamento de F\'{\i}sica, Facultad de Ciencias,
Universidad de La  Serena, Benavente 980, La Serena, Chile}

\slugcomment{{\em The Astrophysical Journal} accepted}

\begin{abstract}
We present a multiwavelength study of the poor cluster RX
J1117.4$+$0743 ([VMF~98]~097) at z$=$0.485, based on
GMOS/Gemini South \sloang, \sloanr~photometry and
spectroscopy, and XMM-Newton observations. We examine its
nature and surroundings by analyzing the projected galaxy 
distribution, the galaxy velocity distribution, the
weak-lensing mass reconstruction, and the X-ray spectroscopy
and imaging. The cluster shows a complex morphology. It is
composed by at least two structures along the line-of-sight,
with velocity dispersions of 592$\pm82$ km s$^{-1}$  and
391$\pm85$ km s$^{-1}$ respectively. Other structures are
also detected in X-ray, in  the galaxy projected number
density map, and by weak-lensing. One of these clumps,
located East from the cluster center, could be
gravitationally bound  and associated to the main cluster.
The derived temperature and bolometric
X-ray  luminosity reveal that [VMF~98]~097 behave like a
normal cluster, in agreement with $L_{X}-T_{X}$ correlation
found for both local ($z=0$) and moderate redshift
($z\sim0.4$) clusters. We find that the mass determination
inferred from weak-lensing is in average 3 to 4.8 times
higher (depending on the model assumed) than the X-ray mass.
We  have two possible explanations for this discrepancy: {\it
i)} the cluster  is in non-equilibrium, then the deviation of
the X-ray  estimated mass from the  true value can be as high
as a factor of two; {\it ii)} the intervening  mass along the
line-of-sight of the cluster is producing an  over-estimation
of the weak-lensing mass. Based on the analysis presented, 
we conclude that [VMF 98]~097 is a perturbed cluster with at
least two  substructures in velocity space and with other
nearby structures at  projected distances of about 1
h$^{-1}_{70}$Mpc. This cluster is an  example of a poor
cluster caught in the process of accreting sub-structures to
become a rich cluster.
\end{abstract}

\keywords{galaxies: clusters: individual: RX J1117.4$+$0743 ([VMF~98]~097) 
- X-rays: galaxies: clusters - gravitational lensing - cosmology: 
observations - dark matter}

\section{Introduction}

Cluster of galaxies are the largest gravitationally bound systems in the
Universe. They are excellent laboratories for studying the large-scale
structure formation, structure mass assembly and galaxy evolution.
Numerical simulations show that massive clusters of galaxies form through
hierarchical merging of smaller structures
\citep[e.g.,][]{West91,Richstone92,Jenkins98,Colberg99}.  Clusters are
complex systems, including a variety of interacting components such as
galaxies, X-ray emitting gas and dark matter. Optical and X-ray studies
show that a large fraction of  clusters contains sub-structures,
revealing that clusters are indeed dynamically active structures,
accreting galaxies and groups of galaxies from their neighborhoods
\citep[e.g.][]{Limaneto03}. Even though  it is thought that rich clusters
form at redshift 0.8--1.2 \citep[or as high as 3.0, see][]{Holden04},
there are numerous evidences (optical, X-ray) that clusters are still
accreting sub-structures at intermediate and low redshifts
\citep[e.g.,][]{vandokkum98,Ferrari05,Gonzalez05}. We may witness the
assembly of rich clusters by observing large groups or poor clusters
which,   in turn, would be the future core of  rich clusters. The
details  of this  process will depend in part on how these large
groups/poor clusters relate  to more nearby structures.

Most galaxies in the Universe are concentrated in low-density
environments (groups and poor clusters). For intermediate
redshifts,  $z\sim 0.3$--0.5,  while massive clusters of galaxies
have been widely studied, the intermediate-mass systems, those
between loose groups and rich clusters of  galaxies, have received
comparatively little attention, either in X-rays or in the optical.
In the context of the hierarchical structure formation  scenario,
an intermediate-mass system is a fundamental player to understand
the process involved in the assembly of massive clusters of
galaxies.

X-rays observations of intermediate-mass structures are
particularly interesting, since the cluster X-ray faint-end
luminosity function has eventually to turn over if the luminosity
function of clusters is to meet  that of single brightest
elliptical galaxies with X-ray luminosities of a  few $10^{41}$
ergs~s$^{-1}$. If this gap at intermediate luminosities could  be
closed, an X-ray luminosity function of all galactic systems could
eventually be established, in analogy to that existing for the
optical \citep{Bahcall79}. In addition, the spatial distribution of
low-mass clusters at intermediate-redshifts could be studied in
order to map regions just entering the non-linear regime, i.e.,
$\delta \rho/\rho \sim 1$.

In the optical, the intermediate-mass systems are also of great
importance and have received little attention. Many previous works
have focused on the study of the galaxy population at intermediate
redshifts but mostly in rich cluster of galaxies. These have
established that the morphological content of galaxy clusters at
intermediate redshift differs dramatically from that in nearby
clusters \citep[e.g.,][]{Dressler97,Oemler97,Smail97}. Indeed, at
$z\sim 0.3$--0.5 there is an excess of spirals and a deficiency of
lenticular galaxies in cluster cores when compared with the galaxy
population in nearby clusters. It has been shown by these studies that
the morphology-density relation is strong for concentrated,
``regular'' clusters, but nearly absent for clusters that are less
concentrated and irregular, in contrast to the situation for
low-redshift clusters, where a strong relation has been found for
both. \citet{Dressler97} suggests that these observations indicate that
the morphological segregation proceeds hierarchically along the time,
i.e. irregular clusters at intermediate redshifts are not old enough
to present segregation. However, nearby irregular clusters seem to be
evolved enough to establish the correlation. Taken together, these
studies reveal that the morphological segregation has evolved
significantly since $z \sim 0.5$, at least for regular clusters.
However, it is not yet known at which degree, if any, morphological
segregation evolves in the sparser environments of groups.

A few poor clusters or groups at intermediate redshifts have been
studied, either in X-ray and/or in the optical
\citep[e.g.,][]{Ramella99,Carlberg01,Wilman05,Mulchaey06}. One example
is the work of \citet{Balogh02}, where it is presented the first
spectroscopic survey of intrinsically low X-ray luminosity clusters
($L_{X} < 4 \times 10^{43}$) at intermediate redshifts $0.23 < z <
0.3$. The ten systems studied have velocity dispersions in the range
350--850 km~s$^{-1}$, and are consistent with the local
$L_{X}$--$\sigma$ correlation. They also find that the spectral and
morphological properties of galaxies in these clusters are similar to
those found in more massive systems at similar redshifts.
More recently, \citet{Jeltema06} described the properties of 6
intermediate redshift groups ($0.2 <z< 0.6$) observed with
XMM-\textit{Newton} and concluded that they follow the same scaling
relation  observed in nearby groups.

In this paper we analyze the properties of the low-luminosity X-ray
cluster of galaxies RX J1117.4$+$0743 -- hereafter [VMF 98]~097 --
based on optical and X-ray data. The cluster was selected from the
160 Square Degree ROSAT Cluster Survey \citep{vik98} and is part of an
ongoing project to study the cluster properties and the galaxy
population of poor clusters in the redshift range $0.15 < z < 0.5$.
This paper is arranged as follows. In Section 2 and 3 we describe the
optical and X-ray data, respectively. Section 4 shows the results
based on the analysis of these data: the velocity distribution, the
galaxy projected distribution, the cluster color-magnitude diagram, a
weak-lensing analysis, and a study of the mass distribution, based on
weak lensing and X-ray emission. In Sect.~5 we discuss the
evolutionary status of [VMF~98]~097 and in section 6 we summarize our 
conclusions. Throughout this paper we adopt when necessary a standard 
cosmological model:
$H_{0}=70\,h_{70}\,$km~s$^{-1}$~Mpc$^{-1}$, $\Omega_{m}=0.3$ and
$\Omega_{\Lambda}=0.7$. At $z=0.485$, 1\arcsec~corresponds to $ 6.0\,
h_{70}^{-1}\,$kpc.

\section{Optical Observations and Data Reduction}

This study is based on data collected with the Gemini Multi-Object
Spectrograph \citep[hereafter GMOS,][]{hoo04} at the Gemini South
Telescope during the system verification process of the instrument.

\subsection{Imaging}

The cluster was imaged through the \sloanr\ and \sloang\ Sloan filters
\citep{fuk96} in 2003 March and May , using the detector array formed by
three $2048 \times 4608$ pixels EEV CCDs. With a pixel size of 13.5
microns and a scale of 0\farcs073 pixel$^{-1}$, the detectors
cover an area of 5.5 arcmin$^{2}$ on the sky. A total of 12 images of 600
sec in \sloanr\ and 7 images of 900 sec in \sloang\ were obtained, giving
an effective exposure time of 7200 seconds and 6300 sec in both filters,
respectively. We adopted a $2\times 2$ binning for the images (0\farcs146
pixel$^{-1}$ on the sky). Offsets between  exposures were used to take
into account the gaps between the CCDs (37 unbinned pixels) and for
cosmic ray removal. All images were observed under good transparency
(photometric) and seeing conditions, with seeing median values of
0\farcs7 and 0\farcs8 in \sloanr\ and \sloang, respectively.

\begin{figure*}[!htb]
\figurenum{1}
\centering
\includegraphics[width=0.87\columnwidth]{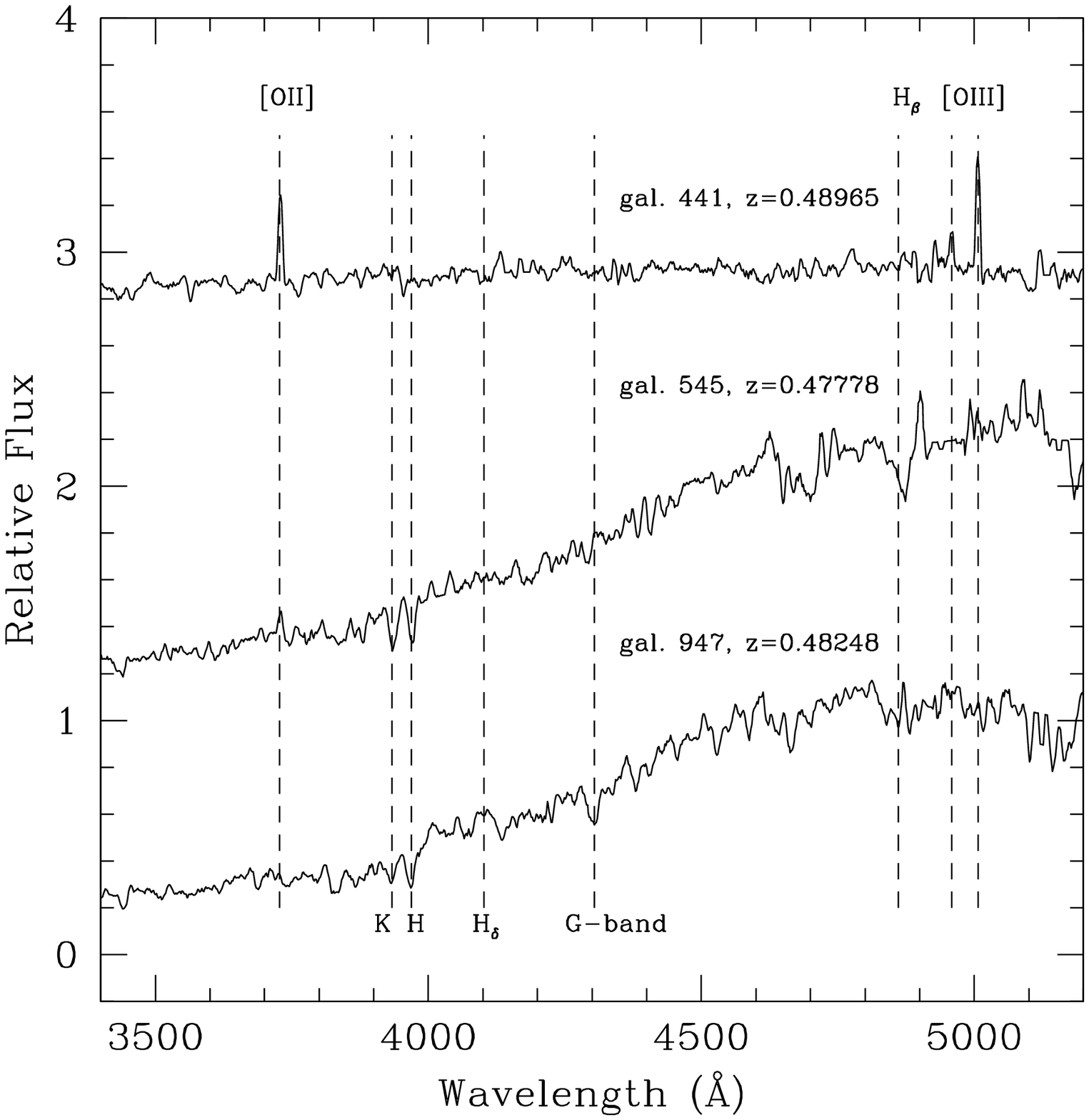}
\includegraphics[width=0.84\columnwidth]{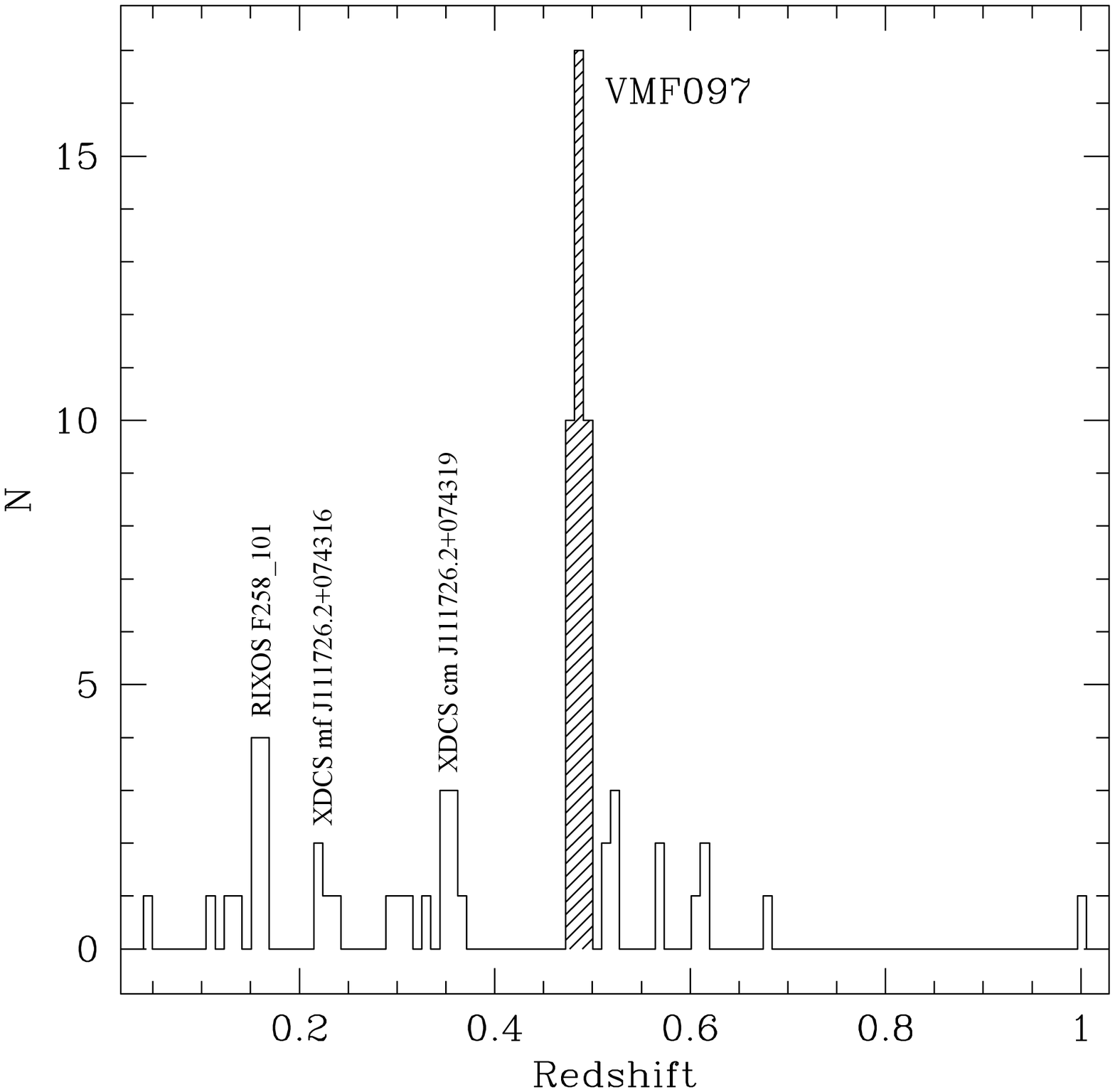}
\caption{{\em Left:} Representative spectra of three galaxy members of
the cluster. From top to bottom: late-type, intermediate-type and early-type.
{\em Right:} Histogram of the redshift distribution of galaxies in the
spectroscopic sample. The shaded histogram, centered at $z \approx 0.485$,
shows the position of [VMF~98]~097 in the redshift space. Other foreground
structures are also shown.}
\label{spec}
\end{figure*}

All observations were processed with the Gemini IRAF package v1.4 inside
IRAF\footnote[1]{IRAF is distributed by NOAO, which is operated by the
Association of Universities for Research in Astronomy Inc., under
cooperative agreement with the National Science Foundation.}. The images
were bias/overscan-subtracted, trimmed and flat-fielded. The final
processed images were registered to a common pixel position and then
combined. The \sloang\ and \sloanr\ magnitude zero-points were derived using
\citet{Landolt92}
standard stars observed immediately before and after the science
exposures. The accuracy of the calibrations is of the order of 5\% and
7\% for \sloanr\ and \sloang, respectively.

We have used SExtractor \citep{ber96} to detect objects in the images and
to obtain their relevant photometric parameters. The combined \sloanr\
image was used to identify objects above a threshold of $1 \sigma_{\rm
sky}$ over the sky level (27.6 mag/arcsec$^2$) and with at least 10
pixels (0.21 arcsec$^{2}$). The photometry in the \sloang-band image was
performed using the parameter ASSOC. This means that the photometric
parameters in \sloang\ were obtained only for those objects detected in
the \sloanr\ image. The resulting catalog was then matched to obtain a
final photometric catalog. We adopted the magnitude given by the
parameter MAG$\_$AUTO as the value for the total magnitude of the
objects. The colors of the objects were determined by measuring their
magnitudes within a fixed 10 pixels diameter circular aperture (1\farcs5),
corresponding to $9.0\, h_{70}^{-1}$ kpc at the cluster rest-frame.

The SExtractor ``stellarity'' index (an indication of how certain an
optical source is unresolved) was used to separate stars from galaxies.
Objects with a stellarity index $\le0.9$ were selected as galaxies. This
cut is in agreement with a separate classification done by plotting pair
of object parameters, like central intensity \textit{vs.} area, central
intensity \textit{vs.} size, and peak intensity \textit{vs.} size , as
well as by visual control. In all cases the classifications are
consistent down to \sloanr=25.5 mag. The galaxy counts calculated using
the objects classified as galaxies reach their maximum at \sloanr~=~25.8
mag. Using this information and the uncertainties in the galaxy
classification above \sloanr~=~25.5 mag, we have adopted this latter
value as our limiting magnitude. The final catalog contains the total
magnitudes, the colors and the structural parameters for 2698 objects
classified as galaxies. Of these, 1348 are brighter than 25.5 mag in \sloanr\
($M_{r\prime}=-16.7$ at the distance of the cluster).

\subsection{Spectroscopy}

The targets for spectroscopic follow up were selected  based on their
magnitudes only. No color selection was applied,  meaning that the sample
includes galaxies of different  morphological types.  All galaxies with
apparent magnitudes brighter than \sloanr$=$23 mag were selected for
spectroscopy (31\% of the total sample).  Of these, only 79 objects were
observed spectroscopically. Two masks were created: one for bright objects
(\sloanr~$\le 20$ mag) and another for faint objects ($20 < $\sloanr~$\le 23$ 
mag).

The spectra of the galaxies were obtained with GMOS in 2003 May 29--30,
during dark time, with a good transparency, and with a seeing that
varied between 0\farcs8 and 0\farcs9. A total exposure times of
3600 seconds and 6000 seconds were used for masks containing bright and
faint objects, respectively. Small offsets of $\sim50$ pixels in the
spectral direction ($\sim 35$\AA) towards the blue and/or the red were
applied between exposures to allow for the gaps between CCDs and to avoid
any lost of important emission/absorption lines present in the spectra.
Spectroscopic dome flats and comparison lamp (CuAr) spectra were taken
after each science exposure. All spectra were acquired using the 400
lines/mm ruling density grating (R400) centered at 6700\AA, in order to
maximize the wavelength coverage for galaxies at the cluster distance.

All science exposures, comparison lamps and spectroscopic flats were bias
subtracted and trimmed. Spectroscopic flats were processed by removing
the calibration unit plus GMOS spectral response and the calibration unit
uneven illumination, normalizing and leaving only the pixel-to-pixel
variations and the fringing. The resulting 2-D spectra were then
wavelength calibrated, corrected by S-shape distortions, sky-subtracted
and extracted to an one-dimensional format using a fixed aperture of
1\farcs3.  The residual values in the wavelength solution for 20--30
points using a 4th or 5th-order Chebyshev polynomial typically yielded
\textit{rms} values of $\sim0.15$--0.20~\AA. With the choice of a
0\farcs75  slit width, the final spectra have a resolution of $\sim
5.5$\AA\ (measured from the arc lines FWHM) with a dispersion of $\sim
1.37$ \AA\ pixel$^{-1}$, covering a wavelength interval of $\sim
4400$--9800~\AA\ (the wavelength coverage depends on the position of the
slit in the GMOS field-of-view). Finally, the residuals of the 5577~\AA,
5890~\AA, and 6300~\AA\ night-sky lines were removed from all spectra
using a 10-th order cubic spline polynomial. Beyond 7800~\AA, the
residuals of night-sky lines were simply masked.

\begin{figure*}[!htb]
\figurenum{2}
\centering
\includegraphics*[width=0.95\columnwidth]{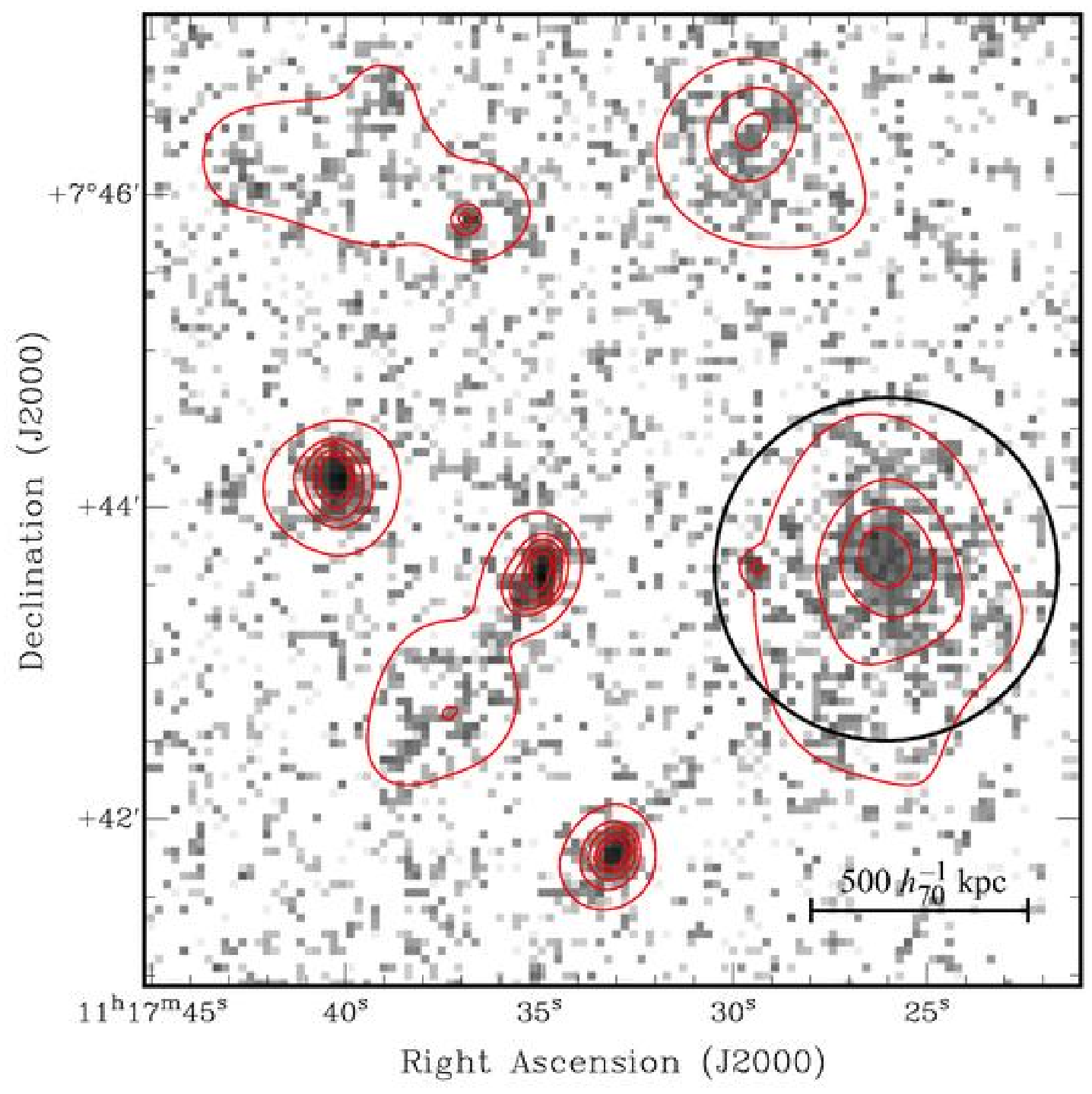}
\includegraphics*[width=0.95\columnwidth]{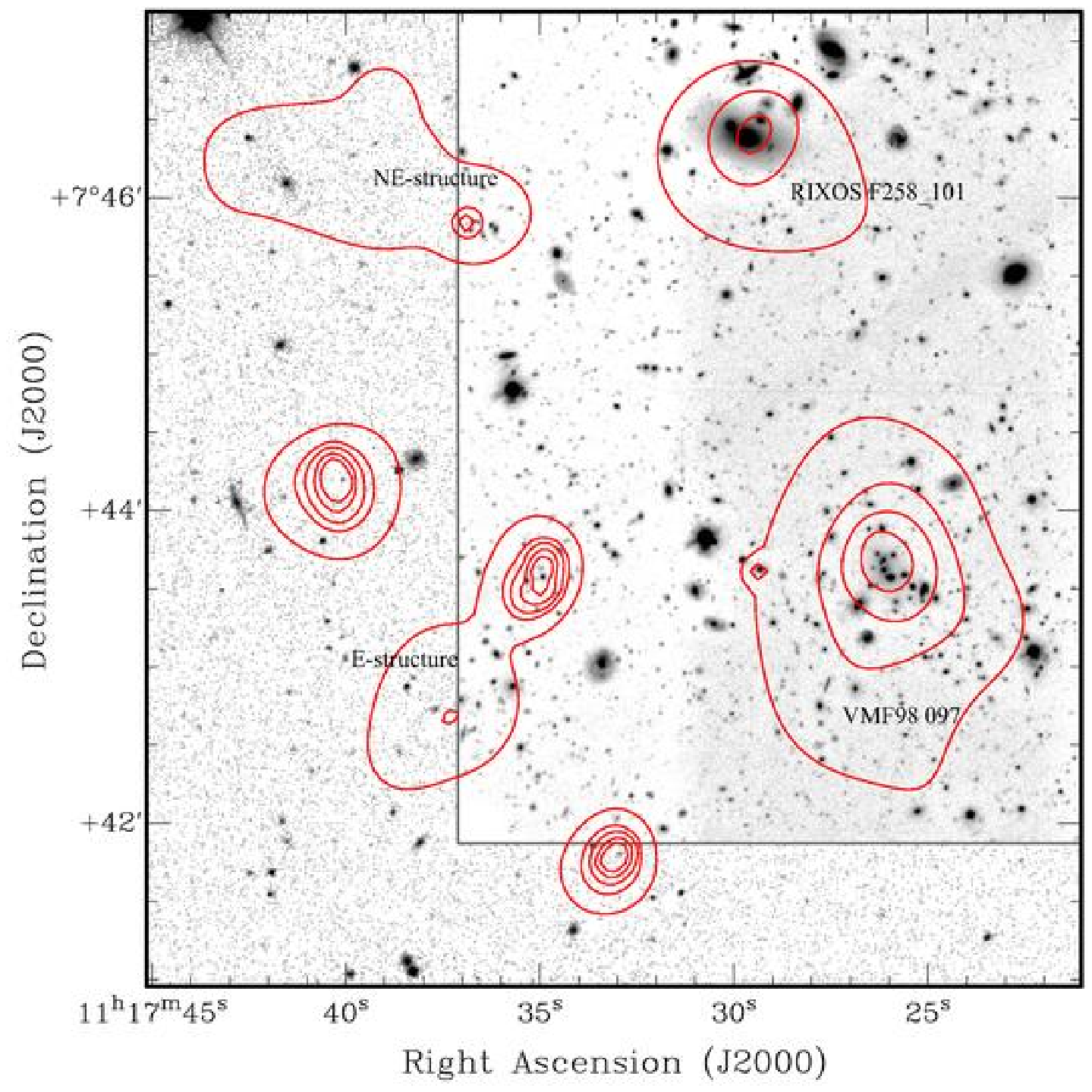}
\caption[]{{\em Left:} Composite image using all EPIC MOS observations where
the cluster is seen entirely, in the 0.3--8.0~keV band. The circle with
66\arcsec radius shows the region where the [VMF~98]~097 spectrum was
extracted (the scale length in the plot assumes $z=0.485$). {\em Right:}
adaptively smoothed X-ray emission plotted over the GMOS \sloanr\ plus SDSS
\sloanr\ composite image. The inner rectangle shows part of the region covered
by the GMOS field. The thin lines in the figures are the isocontours at 0.4,
1.2, 2.0, 3.7, and  $6.3\, \sigma$ above the X-ray background.}
\label{xray_em}
\end{figure*}

To obtain the galaxy radial velocities, we first inspected the spectra to
search for obvious absorption and/or emission features characteristic of
early- and late-type galaxy populations. For galaxies with clear emission
lines, the routine RVIDLINE in the IRAF RV package was used employing a
line-by-line gaussian fit to measure the radial velocity. The residual
of the average velocity shifts of all measurements were used to estimate
the errors. For early-type galaxies, the observed spectra were
cross-correlated with high signal-to-noise templates using the  FXCOR
program in the RV package inside IRAF. The errors given by FXCOR were
estimated using the r statistic of Tonry \& Davis (1979):
$\sigma_{v}=(3/8)(w/(1+r))$, where $w$ is the FWHM of the  correlation peak
and $r$ is the the ratio of the correlation peak height to the  amplitude
of the antisymmetric noise. The left panel in Fig. \ref{spec} shows the
smoothed spectra of three galaxies identified as cluster members,
corresponding to three different spectral types: early-type (bottom),
late-type (top) and  intermediate-type (middle).

We were able to measure redshifts for 77 objects ($\sim 95$\% success rate).
Seventy five of them are galaxies and two are M-class stars. As expected, the
fraction of emission-line galaxies is relatively high and constitute 33\%
of the total sample. However, the fraction of emission-line galaxies that are
cluster members is lower and represent only $\sim22$\% of the
cluster galaxy population. The emission-line galaxy fraction is in agreement
with the  results obtained by \citet{Balogh02} for 10 intrinsically low X-ray 
luminosity cluster ($L_{X}<4$ 10$^{43}$ erg s$^{-1}$) at z$\sim$0.25 
(see section 4.1 for more details). The  measured redshifts, corrected to 
the heliocentric reference frame, and the corresponding errors are listed in
Table \ref{tab1} (columns 6 and 7, respectively). The galaxy identifications
and their sky coordinates are given in the first 3 columns. The apparent
total magnitudes in \sloanr-band and the \sloang $-$ \sloanr\ colors inside a
fixed circular aperture of 1\farcs5 are listed in columns 4 and 5,
respectively. The $R$ value \citep{ton79} listed in column 8 was used as a
reliability factor of the quality of the measured velocity . For $R>3.5$, the
resulting velocity was that associated to the template which produced the
lowest error. For galaxies with $R<3.5$, we looked for absorption features
like CaII and G-band in the spectra, and performed a line-by-line Gaussian
fit using the package RVIDLINE. The resulting values where then compared with
the velocities given by cross-correlation. In all cases the agreement between
the two procedures were good.

The histogram of the redshift distribution is presented in the right panel
of Fig.~\ref{spec}. The concentration of galaxies at $z \approx 0.485$
(shaded area) indicates the position of the [VMF~98]~097 cluster. The peak
at $z \approx 0.16$ corresponds to a group of galaxies, RIXOS F258\_101,
located $\sim$ 2\farcm5 North of the cluster core  (see
Section~\ref{sec:GalProjDist}).  Two other small peaks can be seen which are
probably related to the groups reported by \citet{Gilbank04}.

\section{X-ray Observations and Data Reduction}

The cluster [VMF~98]~097 was serendipitously discovered in X-rays in a
pointed ROSAT PSPC observation of QSO PG1115+080 \citep{vik98}. This object
was observed by XMM-\textit{Newton} in December 2002 (obsID 082340101) and
twice in June 2004 (obsID 203560201 and 203560401). [VMF~98]~097 is found in
the field of view of the MOS1 in all exposures, but it was observed entirely
by the PN detector only in the 2002 observation (only half of the cluster
appears in any MOS2 field of view). This cluster was also observed by
Chandra ACIS-I3 in June 2000 (P.I. G.P. Garmire) in a 26~ks exposure.
However, it produced only $\sim 200$ net counts (background corrected).
Therefore, the Chandra observations are not used in the analysis.

We have downloaded the ODF files from XMM public archives and performed  the
MOS and PN ``pipelines'', which consist in the removal of bad pixels,
electronic noise and correction for charge transfer losses with the program
SAS v6.5.0. We have then applied the standard filters and removed the
observation times with flares using the light-curve of the [8.0--14.0 keV]
energy band. The final exposure times after subtracting high particle
background intervals of the cleaned event files for all observation are given
in Table~\ref{tbl:resumoXMM} together with the net count number (i.e., source
minus background counts).

\begin{figure*}[!htb]
\figurenum{3}
\centering
\includegraphics*[width=0.78\columnwidth]{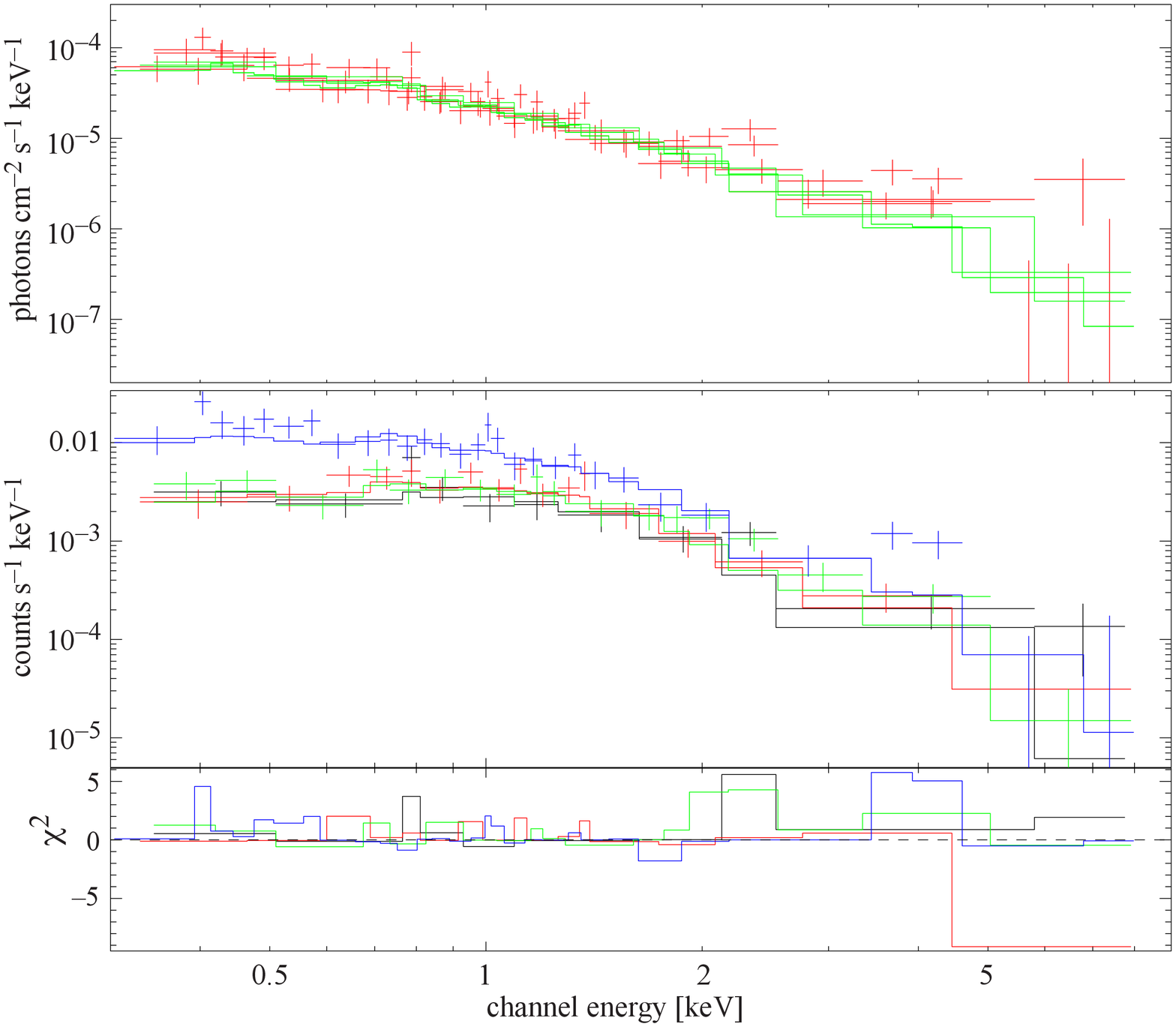}%
\includegraphics*[width=0.90\columnwidth]{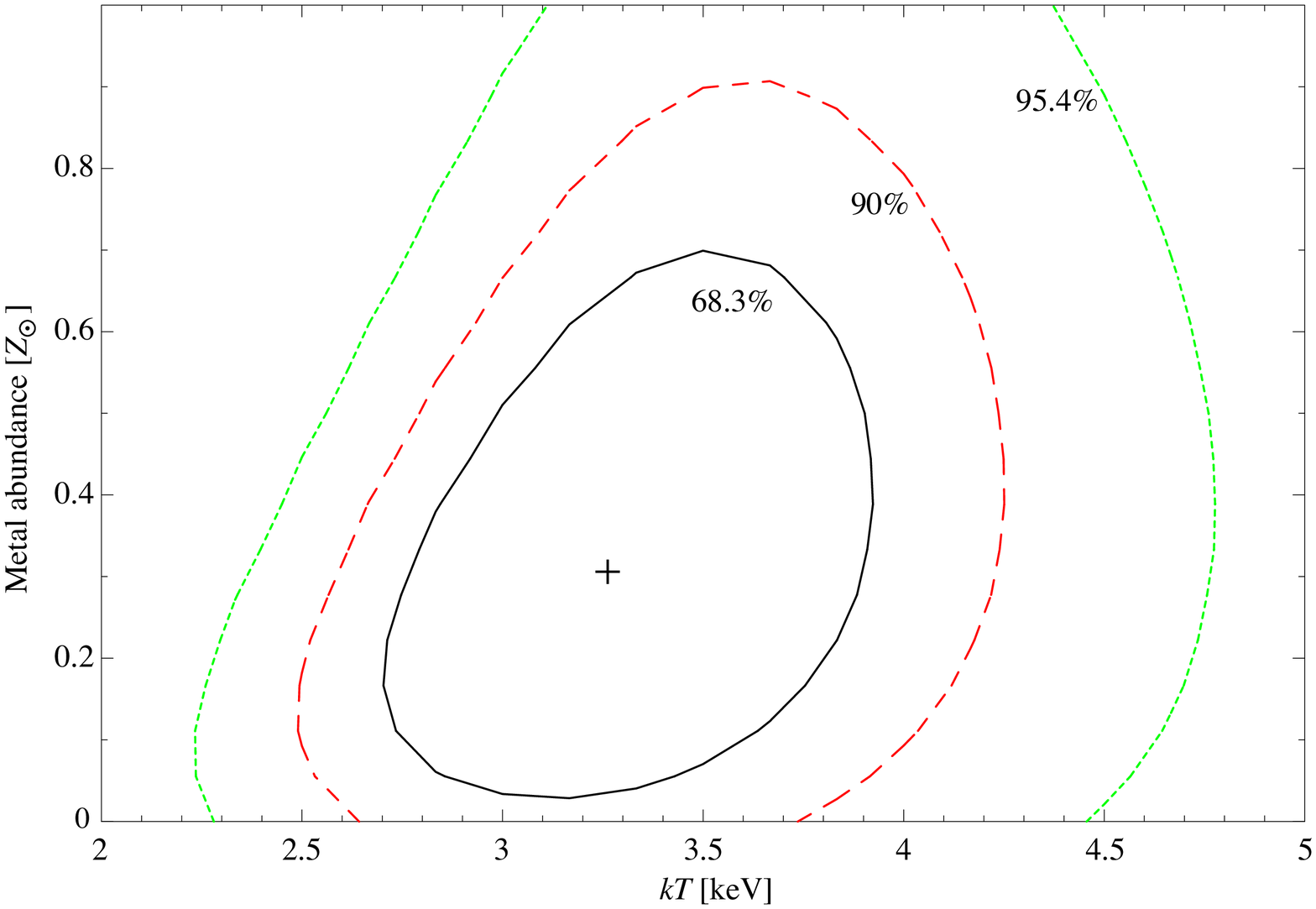}
\caption[]{{\em Left:} Best fit MEKAL model. Top: data from MOS1 and
PN and the best
fit spectrum in units of photons~cm$^{-2}$~s$^{-1}$~keV$^{-1}$, corrected
for instrumental effects; Middle, flux in counts~s$^{-1}$~keV$^{-1}$. The
upper line corresponds to the PN data, the three MOS1 observations are
seen below. Bottom: residuals of the best fit spectrum. {\em Right:}
Correlation between temperature and metallicity obtained with
the spectral fit. The contours correspond to 68.3\%, 90\%, and 95.4\%
confidence levels; the best-fitted value is shown with a plus sign.}
\label{xray_spec}
\end{figure*}

\subsection{X-ray Imaging and Spectroscopy}

For the spectral analysis we have selected a circular region of 1\farcm1
centered at RA~$=11^{\rm h}17^{\rm m}26$\fs$5$, 
DEC~$=+07$\degr$43$\arcmin$33$\arcsec~(J2000). The background was 
selected in the same observation. We have used a larger extraction region 
near the detector border (this because [VMF~98]~097 is itself near the 
border), without any visible sources.

Since about half of the cluster falls outside the MOS2 field of view, we have
used only MOS1 and PN cameras of the 2002 observations, and only the MOS1 of
the 2004 observations. The total source counts, background subtracted, are
also given in Table~\ref{tbl:resumoXMM}. The left panel of Figure
\ref{xray_em} shows the composite image made with MOS1 and PN available data
in the 0.3--8.0 keV band. The EPIC-MOS has a FWHM $\approx5$\arcsec at the
center of the detector. However the MOS point spread function has somewhat
extended wings and the half energy width (HEW) is
$\approx$14\arcsec\footnote[2]{see
\texttt{http://xmm.vilspa.esa.es/external/xmm\_user\_support/\\
documentation/uhb/index.html}}.
For off-axis sources there is a degradation in the resolution, which also
depends on the energy \citep{Ehle06}. The cluster [VMF~98]~097 is located
$\approx$12\farcm5 from the detector axis. Therefore the resolution at 1.5 keV
is 5\farcs5 and 6\farcs5 for the MOS-1 and PN respectively. At 5 keV the
resolution is 6\farcs3 and 7\farcs6 for the MOS-1 and PN respectively. The
effective exposure time is also affected, but it is taken into account by the
Redistribution Matrix File (RMF) and the Auxiliary Response File (ARF)
\citep{Ehle06}.

We also present in the right panel of Fig.~\ref{xray_em} the smoothed  X-ray
emission plotted over the GMOS \sloanr\ image. The core of the cluster, about
1\arcmin~South of the center of the image, is clearly detected in X-rays. The
emission at the North of the image is associated with a foreground group at
$z \sim 0.16$ (see Section~\ref{sec:GalProjDist}). Two other X-ray emission
features are worth mentioning since they are also present in the
galaxy-density map and in the weak-lensing map discussed below: the feature
at the East of the cluster core (hereafter E-structure) and another at the
Northeast, at the border of the optical image (hereafter the NE-structure).

The ancillary and redistribution files (ARF and RMF) were created with
the SAS tasks \texttt{arfgen} and \texttt{rmfgen}, taking into account the
extended nature of the source. The MOS and PN spectra were fitted
simultaneously, each spectrum with its own RMF, ARF, and background files.
The spectral fits were done with XSPEC 11.3 in the range [0.3--8.0 keV]
on the re-binned spectrum, with at least 12 counts per energy bin.

We have used the MEKAL \citep{Kaastra93,Liedahl95} plasma model with a
photoelectric absorption given by \citet{Balucinska92}. For the hydrogen
column density, we have adopted the galactic value at the position of
[VMF~98]~097, $N_{H} = 3.37\times 10^{20}$cm$^{-2}$ 
\citep[using the task \texttt{nh} from \textsc{ftools}, which is an
interpolation from the][galactic $N_{\rm H}$ table]{Dickey90}.

Given the evidences presented in Sect.~\ref{sec:velDist} that the velocity
distribution has two peaks, we have tried the spectral fits with different
redshifts but the results are virtually the same: we cannot, with these
spectra, obtain a redshift estimate of the source. We have also tried a two
component MEKAL model, representing each source in the line-of-sight; however
the fit did not converge because of the low signal-to-noise ratio of the 
spectra. Therefore, we give our results here as  mean emission-weighted values,
adopting a fixed mean redshift.

The best-fit model is shown in the left panel of Fig.~\ref{xray_spec}. Fixing
the redshift at $z = 0.485$ (average redshift obtained for the cluster, see
below), we obtain the following temperature and metal abundance
(metallicity): $kT =3.3_{-0.6}^{+0.7}$~keV and $Z = 0.3_{-0.2}^{+0.4}$
(the errors are at the 90\% confidence levels). The fit is fairly good with
$\chi^{2}$/d.o.f. $= 247.0/250$ (the null hypothesis probability is 0.54).

The metallicity is not well constrained since the Fe-K line is not well
detected. This is also shown in the temperature-metallicity correlation plot,
in the right panel of Fig.~\ref{xray_spec}.

\subsubsection{Measured flux and luminosity}

We have computed the unabsorbed X-ray flux and luminosity in different energy
bands using the plasma model described in the previous section.
Table~\ref{tbl:LX} summarizes these results.

With the determined temperature and bolometric luminosity, [VMF~98]~097 is
found to behave like a normal cluster, in agreement with the local $L_X$--$T_X$
correlation \citep{Willis05} and the correlation for intermediate redshift 
clusters \citep[$z\sim0.4$, ][]{Jeltema06}. The agreement with both local and
intermediate relations comes from the intrinsic scatter in both relations
and the error bars in our cluster. Therefore, the X-ray emission of
[VMF~98]~097 is not affected (at least significantly) by the emission from
the group/structure behind the cluster.

\subsubsection{Radio emission}

\begin{figure}[!htb]
\figurenum{4}
\centering
\includegraphics[width=0.90 \columnwidth]{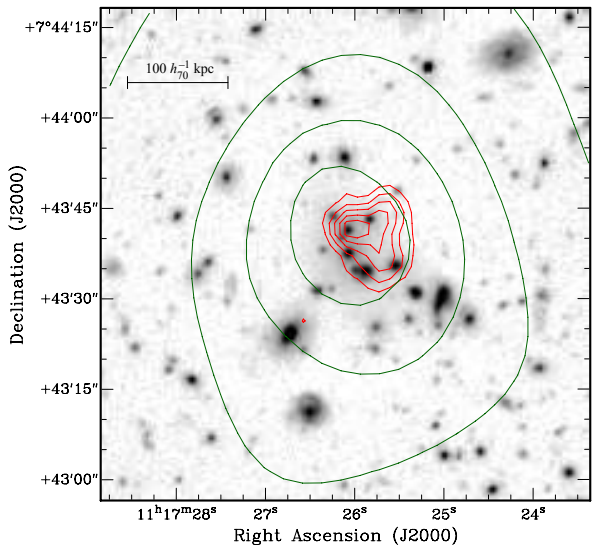}%
\caption[]{GMOS optical image with VLA 1.4~GHz continuum emission
isocontours (central, thick lines) and X-ray emission contours (thin
lines) overlaid.  The radio contours are linearly spaced in intensity. The
X-ray contours are  the same as in Fig.~\ref{xray_em}. Notice the
wide-angle tail morphology of the  radio emission.} \label{fig:vlamosgmos}
\end{figure}

A search with NED\footnote[3]{NASA/IPAC Extragalactic Database,
\texttt{http://nedwww.ipac.caltech.edu/}} reveals that this cluster has a
radio emission at 1.4~GHz associated to it. We have used this information
to look for a radio image in the FIRST survey\footnote[4]{Faint Images of the
Radio Sky at Twenty-centimeters, \texttt{http://sundog.stsci.edu/}}. The radio
image has a FWHM of 5\farcs4 and the radio emission contours are shown in
Fig.~\ref{fig:vlamosgmos}.

The peak of the radio emission coincides with galaxy 1098
(Tab.~\ref{tab1}) and the center of the X-ray emission. It has wide-angle
tail morphology, which is often found in radio galaxies in the center of
clusters \citep{Roettiger96,Gomez97}. This radio morphology implies that
the radio  galaxy is moving with respect to the ICM. This could be due to
the bulk motion  of the intra-cluster gas \citep{Burns02}, for instance,
because of a cluster or group merging. In the present case, based on the
broadening of the radio emission, the direction of motion perpendicular to
the line-of-sight is West--East.

The presence of an AGN may contaminate the X-ray spectrum with a hard
component, making the spectral determined temperature artificially
higher.  The radio emission associated with galaxy 1098 suggests such an
AGN. However, there is no sigh in either XMM or Chandra data suggesting a
point source or an excess X-ray emission at the spatial location of galaxy
1098. Since the X-ray surface brightness is quite flat in the cluster
center, a bright X-ray AGN would be detectable. In addition, there is no
indication of an AGN in the optical spectrum of the galaxy 1098.

\section{Optical Data Analysis}

\subsection{Velocity distribution}\label{sec:velDist}

\begin{figure}[!ht]
\figurenum{5}
\centering
\includegraphics[width=0.95 \columnwidth]{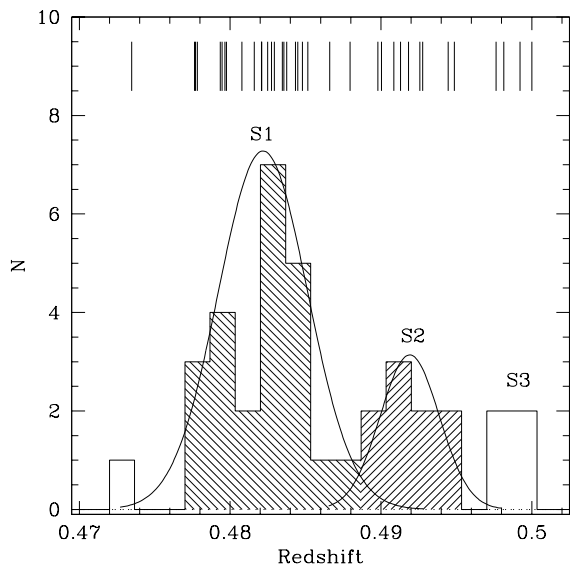}%
\caption{Histogram of the redshift distribution of 37 galaxies in area of
[VMF~98]~097. The two main structures (S1 and S2) are clearly seen in
the plot. A third structure, with 4 galaxies located at $z > 0.492$,
is also shown in the figure.}
\label{hist_v}
\end{figure}

Thirty seven out of 75 galaxies with measured velocities are located in the
redshift interval $0.47 < z < 0.50$, corresponding to the prominent peak seen
in the right panel of Fig.~\ref{spec}. The velocity distribution of these
galaxies is shown in Fig.~\ref{hist_v}. It is clear, from the figure, the
complexity of the cluster. In order to investigate its structure, we use the
KMM test \citep{ash94}, which is appropriate to detect the presence of two or
more components in an observational data set.

First we consider whether the data is consistent with a single component. The
results of applying the test in the homoscedastic mode (common covariance)
yields strong evidence that the redshift distribution of galaxies in the
redshift interval above is at least bimodal, rejecting a single Gaussian model
at a confidence level of 97.7\% (P-value of 0.024). The P-value is another way
to express the statistical significance of the test, and is the probability
that a likelihood test statistic would be at least as large as the observed
value if the null hypothesis (one component in this case) were true. Assuming
two components, they are located at $z=0.482$ and $z=0.494$, corresponding to
the structures S1 and S2 in Figure \ref{hist_v}. They are separated by 3000 km
s$^{-1}$ in the cluster rest frame. 

The histogram shows another gap of $\sim 2000$ km s$^{-1}$ (also in the cluster
rest frame) between S2 and S3 which is formed by 4 galaxies in the interval
$0.496<z<0.510$. If we assume that the velocity  distribution in Figure
\ref{hist_v} is indeed tri-modal, the KMM test rejects a single Gaussian at 
a confidence level of 99\% (P-value of 0.010). Consequently, a model with three
components is statistically more significant than a model with two components.
In this case, the procedure assigns a mean value of $z=0.482$, $z=0.491$ and 
$z=0.498$ with 23 (62\%), 10 (27\%) and 4 (11\%) galaxies for each of the 
structures, respectively.

We used the robust bi-weight estimators $C_{BI}$ and $S_{BI}$ of \citet{bee90}
to calculate a reliable value for the average redshifts (central location) and
the velocity dispersions (scale) of the two main velocity structures (S1 and
S2) present in the cluster. We used an iterative procedure by
calculating the location and scale using the ROSTAT program and applying a
3 $\sigma$ clipping algorithm to the results. We repeated this procedure until
the velocity dispersion converged to a constant value. The best estimates of
the location and scale for S1 and S2 are shown in Table \ref{tab:vel} (columns
5 and 6 respectively). The table also shows, for the velocity structure S1,
the virial radius $R_{v}$ and the virial mass (column 8), computed with the
prescription of \citet{hei85}. The velocity structure S2 is not centrally
concentrated and is probably not virialized (see below). We then chose
to not determine its $R_{v}$ and virial mass. The number of galaxies in the
structure at $z=0.49804$ (S3, see Fig.~\ref{hist_v}) is too small for a
reliable determination of the velocity dispersion and other dynamical
parameters \citep{bee90}.

It is worth noting that the derived line-of-sight velocity dispersion for S1
of 592$\pm$82 km s$^{-1}$ agrees well (inside the 68\% confidence interval) 
with the value inferred from the intra-cluster medium temperature. Indeed, 
using the $T_{X}$--$\sigma$ relation from \citet{Xue00} (which is derived
from  a local sample) the measured X-ray temperature of $kT
=3.3_{-0.6}^{+0.7}$~keV  implies $\sigma = 672_{-53}^{+57}$~km~s$^{-1}$. This
result suggests that the  X-ray emission is associated to S1. This seems to
be the case since the X-ray  emission is centered on the cluster core, which
is associated with the velocity structure S1. 

\subsection{Galaxy projected distribution}\label{sec:GalProjDist}

\begin{figure}[!htb]
\figurenum{6}
\centering
\includegraphics[width=0.95 \columnwidth]{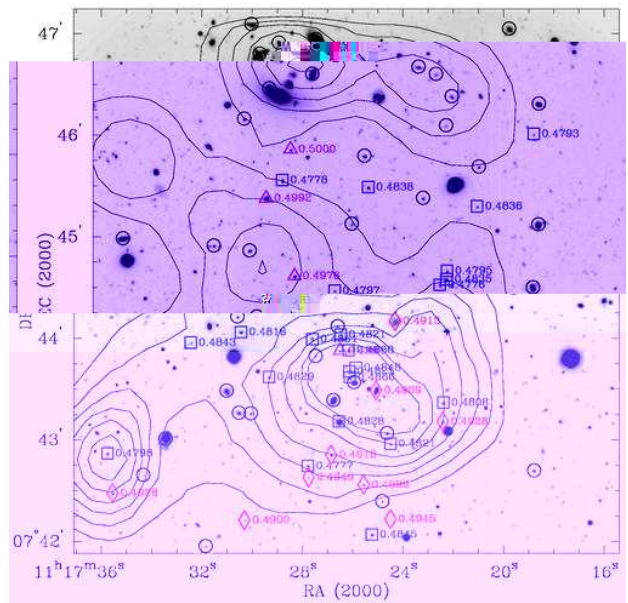}%
\caption{The projected galaxy density map of 272 galaxies brighter than
\sloanr~$=23$ mag. The adaptive kernel density map is superimposed on the
\sloang\ band image ($5.37 \times 5.37\,$arcmin$^2$). The cluster is located
$\sim 1$\arcmin~southward from the center of the image. A second structure,
located towards the NE from the cluster core, seems to be also associated with
[VMF~98]~097. The symbols in the plot represent all galaxies with measured
velocities. The squares and rhombis represent the memeber galaxies of S1 and S2
structures, respectively. The four triangles represent the memeber galaxies of S3. 
The circles indicate the background and foreground galaxies.}
\label{kernel_map}
\end{figure}

Figure \ref{kernel_map} shows an adaptive-kernel density map
\citep[see][]{sil86} based on a sample of 272 galaxies brighter
than \sloanr$=23$ mag. The area corresponds roughly to $\sim 2
\times 2\, h^{-2}_{70}$ Mpc$^2$ at the rest-frame of the cluster.
All structures shown in this map are above the 3 $\sigma$
significance level.

Most of the structures identified in this figure are also
present in the X-ray map (Figure \ref{xray_em}).
The [VMF~98]~097 cluster is represented by the high
density region located  $\sim1$\arcmin~South from the center of the
image. The  second highest density region, located at the top of the
figure, corresponds to a foreground group at
$z \approx 0.16$ \citep[RIXOS F258\_101;][]{mas00}. A third structure, located
$\sim 2$\arcmin~East  from the cluster core, is the E-structure present
in the X-ray map.

Most galaxies in the cluster core have velocities in the range of
S1 (squares), however the galaxies in S2 (rhombi) are mainly distributed,
without any significant concentration, to the South of the cluster core.
We have velocities for only four objects which overlap in space
with the E-structure: one is in S1, another in S2 and two others
corresponding to a nearby and to a background object. The detection
of two velocities at the cluster redshift, as well as the X-ray
emission, suggests that the E-structure is probably dynamically
associated with the cluster.

Figure \ref{kernel_map} shows also two overdensities NE from the cluster
center. The first one, at $\sim 1$\arcmin, may be associated to the cluster
(a substructure), since several velocities in the region are in the redshift 
of the cluster core. The second, at $\sim 2$\arcmin, is the NE-structure in
the X-ray map, and may be either a substructure or a background cluster (the
galaxies there tend to be fainter than those in the cluster core).
Unfortunately we do not have any radial velocity in this region to confirm
this point.

As shown below, the weak-lensing analysis also detect most of the features
present in Figure~\ref{kernel_map}. 

\subsection{The color-magnitude diagram}\label{sec:colmag}

\begin{figure}[!htb]
\figurenum{7}
\centering
\includegraphics[width=0.95 \columnwidth]{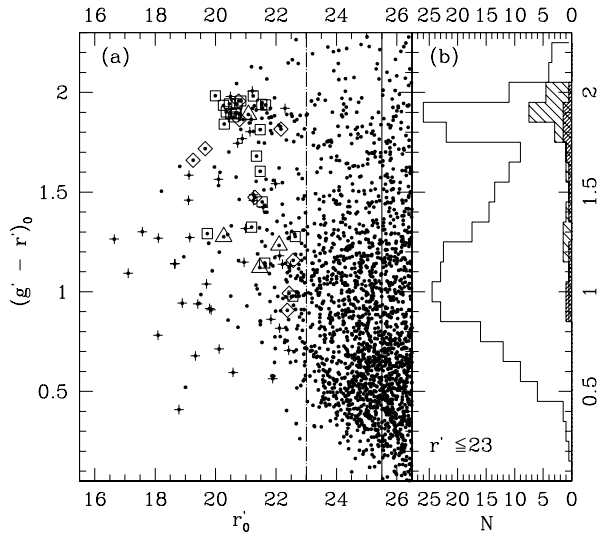}%
\caption{(a) Color-magnitude diagram for all galaxies detected in the
images. Colors and magnitude are corrected for galactic extinction (see
text). The symbols are the same as in Fig.~\ref{kernel_map}
(see Sect.~\ref{sec:GalProjDist}). The solid
line indicates the magnitude of the sample completeness (\sloanr\ = 25.5
mag). The dashed-dotted line represents the limit imposed to select
galaxies for the spectroscopic observations (\sloanr\ = 23). (b) Open
histogram: color distribution of 272 galaxies with \sloanr\ $\le 23$;
Shaded histograms: color distribution of the galaxies members of S1
(shaded to the right) and S2$+$S3 (shaded to the left) structures.}
\label{cmd}
\end{figure}

Galaxy colors provide valuable information about the stellar content of
galaxies, allowing to identify passive and star-forming galaxies in clusters.
Fig. \ref{cmd}(a) shows the color-magnitude diagram (CMD) for all galaxies
detected in the images. Colors and the total magnitudes have been corrected by
galactic extinction from the reddening maps of \citet{sch98} and using the
relations of \citet{car89} ($A_{g}=0.15$ mag and $A_{r}=0.12$ mag,
respectively).

An inspection of this CMD shows that the galaxy populations of the two main
structures of the cluster are not the same, with S1 containing much more red
galaxies (at the so-called cluster red sequence in $(g^\prime-r^\prime)_{0}
\approx 1.9$) than S2 and S3. This behavior is better seen in the color 
distribution histogram of galaxies with measured velocities.
Fig.~\ref{cmd}(b)  indicates that S1 is dominated by a red galaxy population
(right shaded histogram), while S2 and S3 contains mostly blue galaxies (left 
shaded histogram). It is important to note also that S2 and S3 are formed by 
galaxies that, in average, are $\sim 0.3$ fainter than those in S1. 

In a cluster that is dinamically active, it is expected to find a high 
fraction of star-forming galaxies. The results shown above point into that
direction. As we mentioned in section 2.2, the average fraction
of emission-line galaxies in the cluster is relatively small and constitute 
only 22\% of the population. However, there are differences in the
content of the galaxy population in the structures. In S1, only 17\% of the 
galaxies (4 out of 23) are emission-line objects. In the case of S2 and S3, the 
fraction is much higher, and constitute 31\% of the population (5 out of 13). 
The results agree well with what we see in the color-magnitude diagram: the 
emission-line, star-forming (blue) galaxies are more numerous in S2 and S3 than 
in S1. 
 
We investigated the median magnitudes and colors for all galaxies with
\sloanr~$\le 23$ in the two overdensities detected in X-ray, with the
density map (E-structure and NE-structure), and in the cluster core. For
the analysis we select the galaxies inside a radius of 20\arcsec~
($\sim 0.12$ h$^{-1}_{70}$ Mpc at the cluster distance) from the
center given by the maximum of the galaxy overdensity (see
Fig.~\ref{kernel_map}). Table~\ref{tab:mags} summarizes the median magnitudes
and colors for the three overdensities.

The galaxies in the E-structure are $\sim 1$ magnitude fainter and slightly
bluer than the galaxies in the cluster core. The E-structure contains two
galaxies with velocities at the cluster redshift. One of them is a member of
S1, with (\sloang $-$ \sloanr$)=1.92$. The other is a member of S2,
with (\sloang $-$ \sloanr$) = 1.47$, similar to the median color value
derived for this structure. The E-structure is also detected in
X-rays and in the weak-lensing mass map. 

The galaxies in the NE-structure are fainter and bluer than in the cluster
core. This structure is detected in X-ray (Fig.~\ref{xray_em}), and also by
weak lensing (see section \ref{sec:weaklensing}). Due to the lack of redshift
information in this region, we can only speculate about the nature of this
structure, i.e., if it is a background or foreground cluster or even
a sub-structure of [VMF~98]~097.

If the NE-structure is a background cluster, then one would expect a much
fainter galaxy population, but with much redder colors. This is not the case,
since the galaxies in this region are faint, but bluer than in the core of
[VMF~98]~097. Another possibility is that it is indeed a background cluster
of blue, star-forming galaxies, where the red sequence is not yet
established. Finally, this structure could be associated to the foreground
group RIXOS F258\_101 at $z \approx 0.16$  (the NE-structure is located at
1\farcm7 from the center of this group). However this is very unlikely. The
average magnitude of the galaxies in the group is $\sim 18.8$ mag with a
median color of (\sloang $-$ \sloanr$) =1.2$. Although the median color
obtained for the NE-structure is similar to the median color of the galaxies
in RIXOS F258\_101, the galaxy population is much fainter (3.5 mag).

\subsection{Weak Gravitational Lensing Analysis}

Gravitational lensing is a powerful tool for studying the matter
distribution in galaxy clusters. In its weak regime gravitational
lensing allows the reconstruction of the projected mass distribution
through the analysis of the small morphological distortions induced
by gravitational lensing of background sources (weak shear field).
This technique is completely independent of the dynamical state of
the cluster. In this section we apply a weak-lensing analysis to the
imaging data to estimate the mass distribution on the field of [VMF
98]~097.

\subsubsection{Galaxy shape measurements}

The determination of the shapes of faint, putative background
galaxies, was performed using the method described in
\citet{cyp04,cyp05}. In the following paragraphs we summarize the
main steps of the procedure used in the analysis.

We performed galaxy shape measurements, including the removal of
seeing effects and PSF anisotropies, using the algorithm {\sc
im2shape} \citep{bri02}. This program models an astronomical object
as a sum of Gaussian functions with an elliptical base and carries
out the deconvolution of the object image with a local PSF extracted
from the image itself. While stars were modeled as one simple
Gaussian, galaxies are treated as a sum of two Gaussians with same
ellipticity and position angle.

\begin{figure*}[!htb]
\figurenum{8}
\centering
\includegraphics[width=0.88\columnwidth]{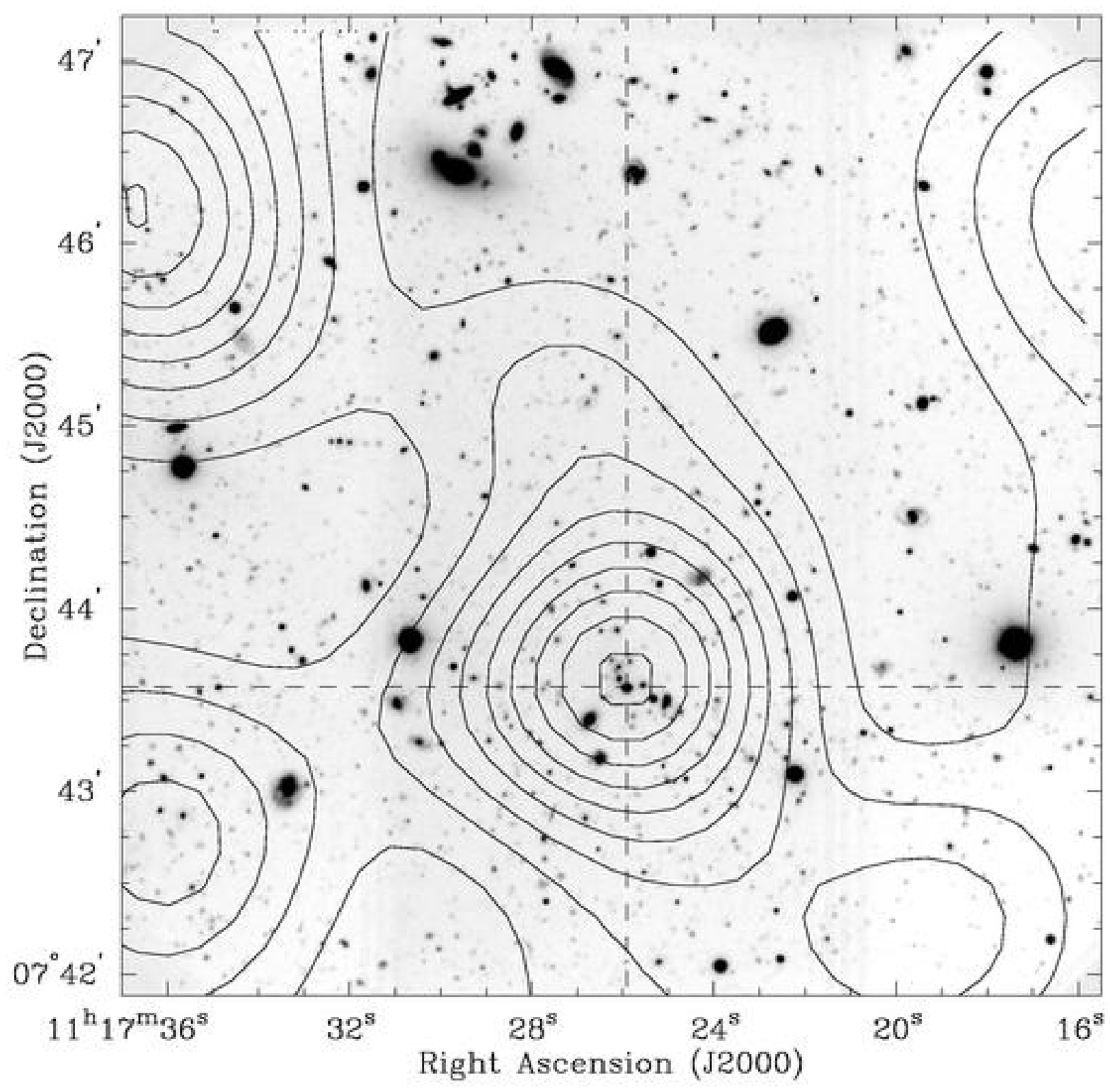}
\includegraphics[width=0.88\columnwidth]{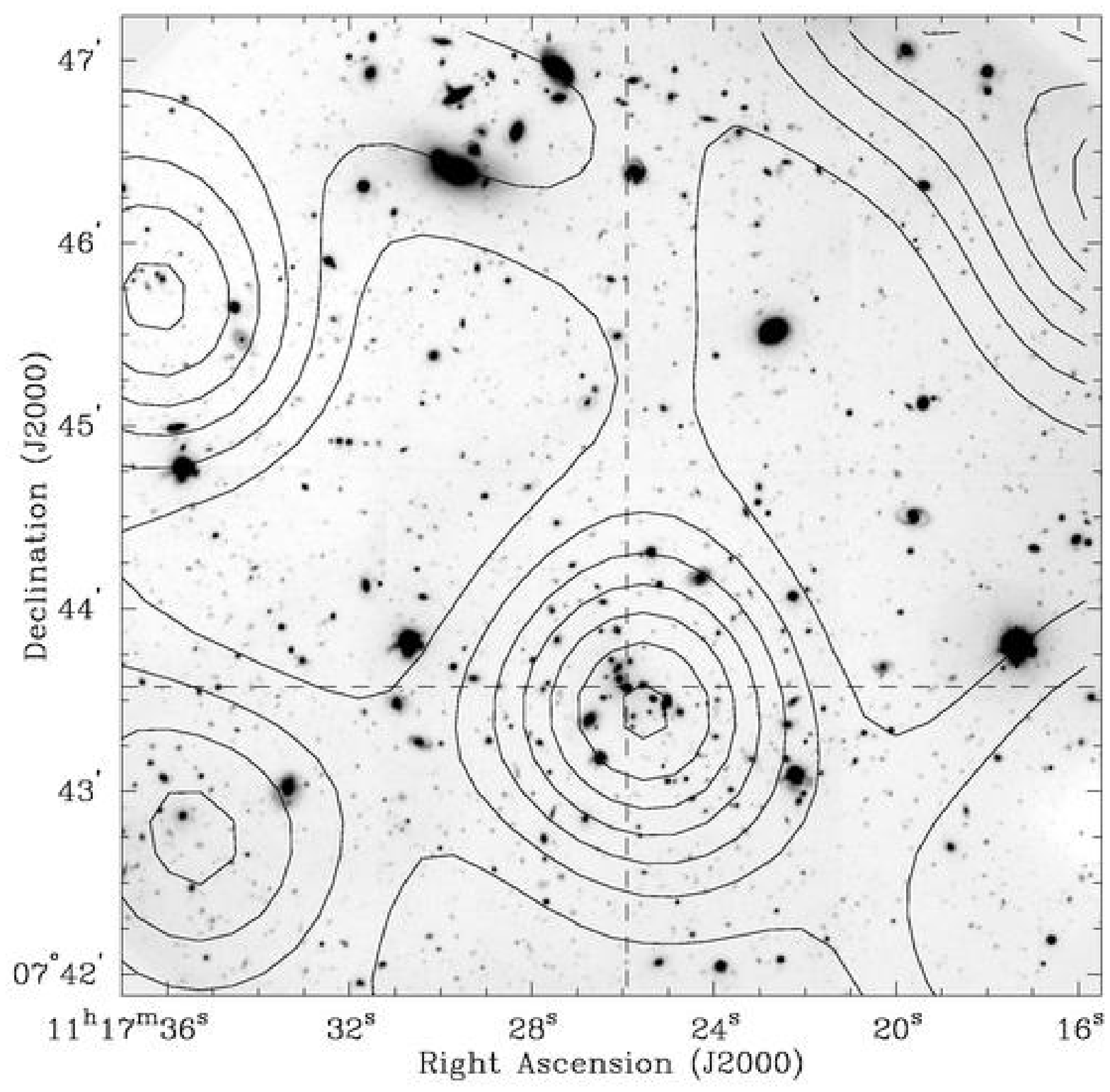}
\caption{Weak-Lensing reconstructed projected mass distributions. {\em Left
panel:} Optical images of [VMF~98]~097 superimposed by constant mass density
contours derived from weak shear measurements done over the \sloang\ image.
{\em Right panel:} mass density contours derived from \sloanr\ image. The
contour levels starts at 4\% of $\Sigma_{\rm crit}$ (the lensing critical
density; $3\times 10^3M_\Sun$ pc$^{-2}$ in the present case), and have linear
steps of 2\%.}
\label{massmaps}
\end{figure*}

We use high signal-to-noise unsaturated stars ($21.5<r^\prime<24.5$)
to map the PSF all over the frame. To make the final catalog,
stellar objects with discrepant ellipticity or full width at half
maximum (FWHM) were removed through a sigma-clipping procedure. In
both images the PSF showed to be nearly constant across the entire
field, having an average ellipticity of 4.4\% and 6.0\% in the
\sloang\ and \sloanr\ images, respectively.

To select background galaxies, which are the probes of the weak shear 
field, we need to rely on their magnitudes and colors to discriminate 
them from the cluster and/or foreground objects, given that we have 
not redshift information for the vast majority of them. Ideally we
would like to use just galaxies redder that the cluster red-sequence
for those objects are supposed to be all behind the cluster
\citep[e.g.][]{Broadhurst} but unfortunately, as we can see in
Fig.~\ref{cmd}, the red-sequence at $z\sim0.5$ is very red for this
particular combination of filters, and the number of galaxies redder
than the red-sequence is too small to provide an adequate sample.

We opt therefore for a simple magnitude and signal-to-noise cut
defining the weak lensing sample as all galaxies fainter than
\sloanr$=$23.0 mag ($M_r = -19.2$ at $z=0.485$) with ellipticities
measured with precision greater than 0.2. This criteria left us with
a sample of 1001 (23 gal.~arcmin$^{-2}$) and 1298 (30 gal.
arcmin$^{-2}$) galaxies for the \sloang\ and \sloanr\ images,
respectively with an average magnitude of \sloanr$=$24.9 mag. By
using this criteria we expect some contamination by cluster or
foreground galaxies to be present but it should not introduce any
bias in the mass reconstruction, only increase the noise.

\subsubsection{Surface mass density distribution}

The surface mass distribution of the cluster [VMF~98]~097 has been recovered
from the shear data using the second version of the {\sc LensEnt} code
\citep{bri98,mar02}. This algorithm takes the shape of every galaxy image as
an independent estimator of the local reduced shear field. The reconstruction
of the mass distribution incorporates an intrinsic smoothing whose
characteristic scale is determined by Bayesian methods, using a maximum
entropy prior. This scale is chosen by maximizing the evidence, given the
input data. Using a Gaussian function to smooth the data, we found that its
optimal FWHM is 70\arcsec.

\begin{figure*}[!htb]
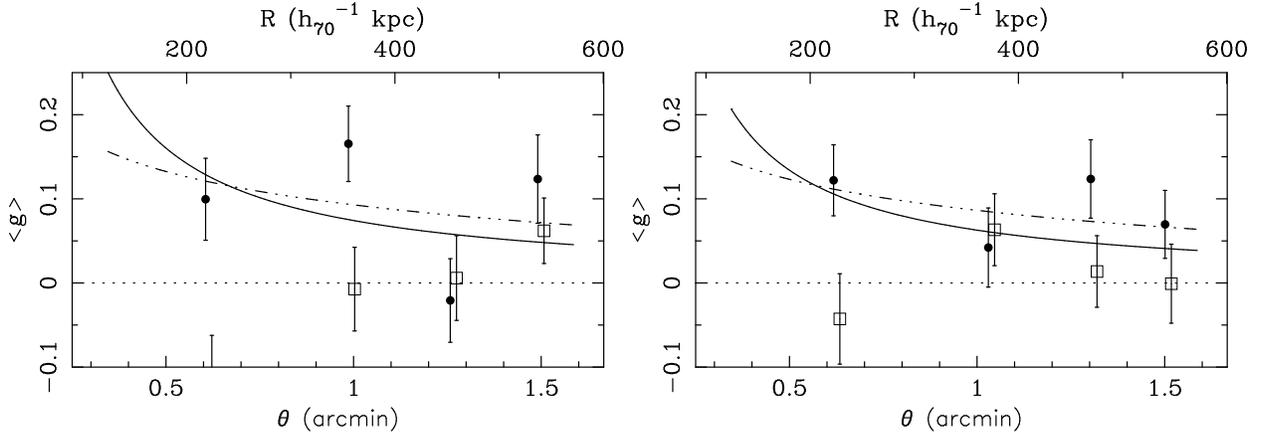

\figurenum{9}
\centering
\includegraphics[width=0.95\columnwidth]{f9a.eps}
\includegraphics[width=0.95\columnwidth]{f9b.eps}
\caption{
Radial profile of the reduced shear $\langle g\rangle$ for the \sloang\ (left
panel) and \sloanr\ (right panel). Filled circles represent the average
ellipticities of background galaxies, projected tangentially with respect to
the cluster center. Each point contains  one third of the galaxies
($\sim80$). The error bars are $1\sigma$ errors of the mean. Empty squares
are the same but for ellipticities projected in a direction 45\degr\ with
respect to the tangential direction, and should average zero. The solid and
the dot-dashed lines represent the best fit of  SIS and NFW profiles,
respectively.}
\label{shear_profile}
\end{figure*}

Figure \ref{massmaps} shows the maps of the reconstructed surface mass
density obtained by using data from each of the images. The maps are very
consistent with each other, all showing basically the same features: a main
structure clearly associated with the core of [VMF~98]~097, and two smaller
structures at the Eastern edge of the field, which can be associated with the
E- and NE-structures discussed above. In the \sloang\ map there is a hint of
a substructure between the cluster core and the NE-structure. The maps also
suggest that there may exist a mass filament joining the E-structure with the
main core. In general, it is actually impressive how X-ray emission,
surface-mass and galaxy-number densities compare well in this field.

Figure \ref{massmaps} also shows the mass center adopted for the radial analysis
presented below (dashed line). This center correspond to one of the
brightest red galaxies on the cluster core which is close to the peak of both
mass maps, particularly the one reconstructed with the \sloang\ image.

\subsection{Mass determination}

We now address the measurement of the cluster mass, considering estimates
based on weak-lensing of background galaxies and on the ICM X-ray
emission.

\subsubsection{Weak lensing}\label{sec:weaklensing}

For mass estimation through weak-lensing, we opt to use physically motivated
mass-density models, to avoid the mass-sheet degeneracy bias \citep{gor88}.
The two most widely adopted models for fitting shear data are the singular
isothermal sphere (SIS) and the NFW profile. The first is a solution of the
hydrostatic equilibrium equation for an isothermal self-gravitating system,
whereas the second provides a good fit to dark matter halos in numerical
simulations \citep{NFW}.

The SIS profile has the advantage of having a single parameter, $\sigma$,
which is associated with the line-of-sight velocity dispersion of the
galaxies. This density profile is given by:
\begin{equation}
\rho(r) = {\sigma_{cl}^2 \over 2 \pi G r^2}
\end{equation}
The NFW profile is described by:
\begin{equation}
\rho(r) = {\rho_{c}~\delta_c \over \left(r /r_s \right)
   \left({1 + r / r_s}\right)^2}
\end{equation}
where $\rho_{c}$ is the critical density, $r_s$ is a scale radius and
$\delta_c$ is given by
\begin{equation}
\delta_c = {200 \over 3} {c^3 \over \ln{(1+c)} + c /(1+c)}
\end{equation}
where $c$ is the  concentration parameter. The approximate virial radius
$R_{200}$ can be defined as $c \times r_s$.
Lensing formula for the SIS and NFW profiles came from \citet{king01}.

These parametric models were fitted to the peak of the mass map corresponding
to the cluster core using the procedure described in \citet{cyp04}. We
restricted the data to the region 15\arcsec $< r <$ 1\farcm6 in relation to
the mass center showed in Fig.~\ref{massmaps}. Data points closer to the
center have been removed because they correspond to a region where the lensing
effects are no longer linear (strong lensing region) and 1\farcm6 is the
distance to the closest image border. Given these limits, the number of data
points included in this analysis is 302 and 373 for the \sloang\ and \sloanr\
images, respectively.

The reduced shear (and the derived mass profile parameters) depends on the
mean redshift of the background galaxies through the mean value of the ratio
$\beta \equiv D_{ls}/D_s$ of the angular diameter distances between the
cluster and the sources and to the sources. We have estimated this quantity
for our sample of background galaxies using a catalog of magnitudes and
redshifts in the Hubble Deep Field \citep{HDF} with both the same bright limit
cutoff and the same average magnitude, obtaining in both cases
$\langle\beta\rangle=0.44$.

In Fig.~\ref{shear_profile} we plot the binned data points of the galaxy
ellipticities as well as the best fitted SIS and NFW models. The best fitted
parameters of these models are presented in Table \ref{weak_results}.

Our data poorly constrains the NFW concentration parameter because it controls
the variation of the density slope in the very inner ($r \ll r_s$) or outer
($r>R_{200}$) regions, which we do not probe in our weak-lensing analysis.
Consequently, we decided to keep the value of $c$ constant, $c=5$, and fit
only $R_{200}$.

The results obtained for the \sloanr\ and \sloang\ images are fully consistent
within the errors. The same is not valid for the two models. The SIS results
tend to give smaller values for the cluster mass when compared with the
results obtained with the NFW model, because within the restricted radial
range we are considering here the SIS profile is steeper than the NFW.

It is worth mentioning that the values of $R_{200}$ obtained through
weak-lensing are significantly above those from the virial analysis presented
in Sec.~\ref{sec:velDist}. We shall come back to this issue in Sec. 5.

\subsubsection{X-ray Brightness profiles and mass determination}

The gas-density profile is obtained from the radial X-ray
surface-brightness profile. We assume that the gas has a number-density
profile given by the $\beta$-model \citep{Cavaliere76}:
\begin{equation}
    n(r) = n_{0} \left[1 + \left(r/r_{c}\right)^{2}\right]^{-3\beta/2}
    \, ,
    \label{eq:gasbaeta}
\end{equation}
where $r_{c}$ is the core radius and $n_{0}$ the central number-density of
electrons. Then, the X-ray surface-brightness profile is:
\begin{equation}
    I(R) = I_{0}  \left[1 + \left(R/R_{c}\right)^{2}\right]^{-3\beta + 1/2}
    \, ,
    \label{eq:SBX}
\end{equation}
assuming that the gas is isothermal and the core radius $R_{c} = r_{c}$
(capital and lowercase symbols refer to 3D and projected quantities,
respectively).

The X-ray brightness profile of [VMF~98]~097 was obtained using the STSDAS/IRAF
task \textsc{ellipse}, with the sum of the MOS1 images corresponding to
observations with obsID 203560201 and 203560401 in the 0.3--8.0~keV energy
band. Each image was binned by a factor 128 so that 1 image pixel was
6\farcs4. The brightness profile was extracted up to 125\arcsec,
corresponding to $\sim 750 ~ h_{70}^{-1}\,$kpc at the cluster
redshift.

\begin{figure}[!htb]
\figurenum{10}
\centering
\includegraphics*[width=\columnwidth]{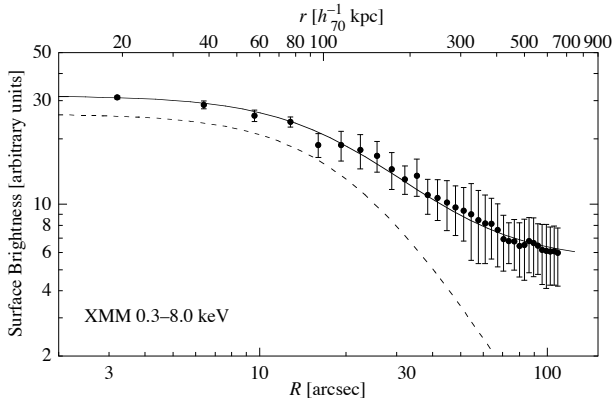}
\caption[]{X-ray surface brightness profile in the [0.3--8.0 keV] band
fitted by a $\beta$-model. The full line is the fit to the cluster plus
background emission. The dashed line corresponds to the
inferred X-ray emissivity of the cluster.}
\label{x_surf}
\end{figure}

Figure~\ref{x_surf} shows the cluster core X-ray emission profile together
with the best least-squares fitted $\beta$-model. We have obtained $r_{c}=$
21\farcs1$\pm$0\farcs9 ($127 \pm 5\,$kpc) and $\beta = 0.526 \pm 0.014$. The
central electronic density, $n_{0}$, is estimated using the emission integral,
$EI = \int n^{2}_{e} dV$ \citep[see][]{Sarazin88}, which is related to the
thermal spectrum normalization parameter given by XSPEC. Using the thermal
spectrum extracted within 66\arcsec, we have obtained $n_{0} = (3.4 \pm 0.4)
\times 10^{-3}\,$cm$^{-3}$.

\begin{figure}[!htb]
\figurenum{11}
\centering
\includegraphics*[width=\columnwidth]{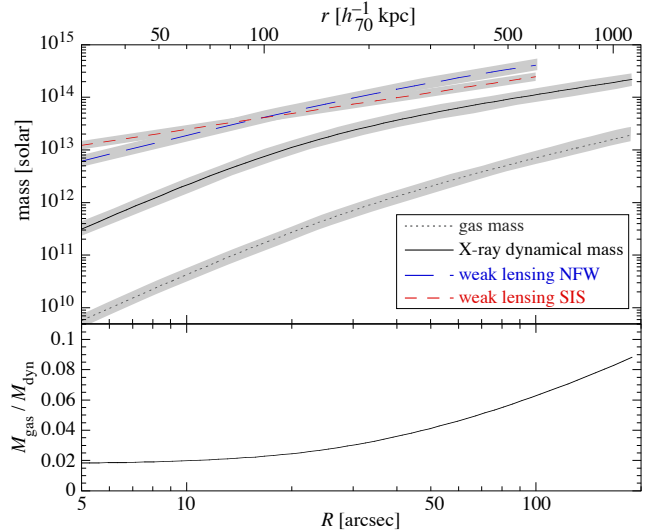}
\caption[]{{\em Top panel:} Gas (dotted line) and X-ray dynamical mass
(solid line) growth curves derived from the assumed $\beta$-model. They are compared 
to the weak-lensing mass estimated with a SIS (short-dashed line) and NFW 
(long-dashed line) profiles. The grey regions correspond to $1 \sigma$ statistical errors.
{\em Bottom panel:} Gas mass fraction profile derived using the X-ray data.}
\label{x_wl_mass}
\end{figure}

The radial gas-mass profile can be simply obtained by integrating the density
profile in concentric spherical shells.
The dynamical (total) mass is computed assuming an isothermal gas in
hydrostatic equilibrium. Even summing all XMM observations, we have enough
counts only to compute a single emission-weighted temperature. Using the
temperature previously determined, $kT = 3.3\,$keV, the computed dynamical
mass is presented in Fig.~\ref{x_wl_mass}. At  $r = $125\arcsec, the total
mass inferred from X-rays is $1.4 \times 10^{14} M_{\odot}$.

The gas mass fraction, $f_{\rm gas}$, is computed as the ratio between the gas
mass and the total mass at a given radius. This ratio is related to the
cluster baryon fraction as $f_{\rm baryon} = f_{\rm gas} (1 + M_{\rm
gal}/M_{\rm gas})$, where $M_{\rm gal}$ is the baryonic mass in the galaxy
cluster members. The baryonic mass in galaxies may be estimated as $M_{\rm
gal} \approx 0.16~h_{70}^{0.5} M_{\rm gas}$ \citep{White93,Fukugita98}. The
bottom panel of Fig.~\ref{x_wl_mass} shows the baryon fraction radial profile.
At $r = $125\arcsec, $f_{\rm gas} = 0.07$, with a rising trend. The X-ray
observations are not deep enough to detect the point where $f_{\rm gas}$
flattens, as is observed in several clusters \citep[e.g.,][]{Allen02}.

\subsubsection{A comparison between weak-lensing and X-ray masses}

At a radius of 0.5 h$^{-1}_{70}$ Mpc, the inferred week-lensing masses 
from the \sloanr\ image are $2.1 \times 10^{14}M_\odot$ and $3.4\times
10^{14}M_\odot$  for the SIS and NFW profiles, respectively. At the same
radius, the X-ray mass is $7.0 \times 10^{13}M_\odot$, i.e. the
weak-lensing mass is 3.4 (SIS) to 4.8 (NFW) times the
value inferred through the X-ray emission. The possible causes
for such discrepancy are given in the next section.

\section{Discussion}

Our results strongly suggest that we are witnessing the mass assembly of a
cluster at $z=0.485$. There are several hints pointing towards this
suggestion.

The morphology of the cluster is complex, presenting at least two significant
substructures. The X-ray emission, the galaxy distribution, and the
surface-mass density map, all present the same overall features: the cluster
core and the E and NE-structures. The E-structure and the cluster core have a
projected distance less than 1 Mpc (at the cluster redshift) and, as we
argued in sections 4.2 and 4.3, are probably at the same redshift. The
substructure located between the cluster core and the NE-structure seem in
the projected galaxy distribution (Figure  \ref{kernel_map}) might be also
real, since there is a feature present in our \sloang\ weak-lensing map close
to it.

The velocity distribution of the region is also complex, with multiple peaks
in a small redshift range (Fig.~\ref{hist_v}). The statistical analysis
presented in Section 4.1 indicates, for instance, the presence of three 
velocity-substructures: S1, S2 and S3.

Additional evidence of dynamical activity in clusters can be obtained
from its brightest galaxy. For local X-ray luminous groups and poor
clusters, the most luminous galaxies (BGGs) lie near the peak of the
X-ray emission  \citep[e.g. ][]{Mulchaey98}. However, at intermediate
redshift, this picture  may be different. A recent study of X-ray 
groups and poor cluster at  moderate redshif by \citet{Mulchaey06}
suggest that the brightest galaxies in groups and poor clusters are
still in the  process of forming, as late as at $z\sim$0.2, in some
systems. The indication is given by the offset  between the BGG and the
X-ray emission and offsets between the velocity of  the BGGs and the
mean velocity of the system. This scenario is consistent  with recent
numerical simulations \citep{deLucia07}. In the case of [VMF~98]097, the
X-ray emission is associated with the velocity structure S1. This
structure has two elliptical galaxies in the center with comparable
luminosities: galaxy 947 with \sloanr$=$20.26 mag, $cz=145882$ km
s$^{-1}$,  and galaxy 1098 with \sloanr$=$20.29 mag, $cz=144645$ km
s$^{-1}$. The peak of the X-ray emission is offset by
$\sim9$\arcsec~from both galaxies.  Furthemore, the brightest elliptical
galaxy (number 947) is offset  significantly in velocity from the mean
velocity of the structure S1 ($>$ 1000 km s$^{-1}$). These results
provide additional evidences of an on-going dynamical activity in this 
cluster.

The mass estimates obtaining using X-rays and weak-lensing are very
discrepant, and this is usually interpreted as evidence of dynamical
activity \citep[e.g.,][]{cyp04}. In fact, X-ray mass estimates are based on
the assumption of hydrostatic equilibrium, which may not hold in the
presence of mergers or strong tidal interactions. It also depends on the
assumption that the $\beta$-model describes well the gas density radial
profile.

The accuracy of mass determination based on X-ray observations has been
studied through hydrodynamical simulations. \citet{Evrard96} and
\citet{Schindler96} find that, usually, X-ray estimates are good within 50\%,
with no systematic bias. In the case of an extreme non-equilibrium clusters,
the mass deviation from the true value can be as high as a factor 2. However,
more recently, \citet{Rasia06} conclude that X-ray estimation of the total
mass is biased towards lower values when using the $\beta$-model if the
cluster is not in equilibrium. The typical estimated mass is around 40\% of
the true mass at  about half the virial radius.

Another possibility is that the excess between lensing and X-ray masses (a
factor $\sim 3.6$) is due to the intervening mass along the line-of-sight of
the cluster, which leads to an over-estimate of the weak-lensing mass of a
few tens of percent \citep{Metzler01}, although this excess seems too high to
be due to this effect only. On the other hand, the velocity dispersion in the
region of the cluster core is consistent with equilibrium between gas and
galaxies (Sect.~\ref{sec:velDist}).

Based on the color-magnitude analysis presented in Sect.\ref{sec:colmag}, it
is unlikely that the E-structure is a foreground group. Besides, the
E-structure contains two galaxies at the cluster redshift and is detected by
weak-lensing mass reconstruction and by X-rays. It lies roughly at 1
$h_{70}^{-1}$~Mpc from the center of [VMF~98]~097 and thus could be a
substructure of the cluster.

The NE-structure poses a more challenging problem, since we do not have any
galaxy with measured redshift in this region. Nevertheless, it is an
overdensity detected in X-rays, in the galaxy projected number density map,
and also by weak lensing. This suggests that it should be relatively
massive. In order to probe its nature, we have compared their
median magnitudes and the color (\sloang$-$\sloanr) with those of the
cluster core and the E-structure. Galaxies in the  NE-structure are
significantly fainter and bluer and, consequently, we suggest that it
is indeed a background cluster, although spectroscopic information is
necessary to verify such a claim.

\section{Summary and Conclusion}

We have presented an optical and X-ray based study of [VMF~98]~097 (RX
J1117.4$+$0743), an intermediate mass structure located at z$=$0.485. We
demonstrate in this work that it is possible to obtain a good
weak-lensing data for a distant cluster even with non-exceptional seeing
conditions (0\farcs7--0\farcs8) and with small field of view. Our main
results are summarized in the next paragraphs.

The cluster shows a very complex structure. We find that velocity
distribution of member galaxies is at least bimodal,  with two well
defined structures along the line-of-sight. The two main structures, 
named S1 with 23 galaxies and S2 with 9 galaxies, form the cluster  core.
These structures have a velocity dispersion of 592$\pm$82 km s$^{-1}$
and 391$\pm$85 km s$^{-1}$  respectively. Using the projected density
map of  272 galaxies brighter than \sloanr$=$23 mag we were able to
identify several structures in the neighborhood of [VMF~98]~097. These
structure are also presented in X-ray and in the weak-lensing maps. The
high density  regions identified in the maps are: the cluster core
formed by S1 and S2, the foreground group  RIXOS F258\_101  (located
$\sim$ 2\farcm5 North from [VMF~98]~097), and two other  overdensities,
the E-structure and the NE-structure. We do not have  redshift
information of the galaxies belonging to these two structures, except
for two galaxies at the cluster redshift in the E-structure. Therefore,
we have used the median magnitudes and colors of the galaxies inside a
20\arcsec~radius from the peak of these overdensities to investigate the
possibility if these structures are linked to [VMF~98]~097. Based on
this analysis and in the detection given by the weak-lensing and X-ray
maps (section 4.3) we conclude that the E-structure is a sub-structure
associated to the cluster. Using the same approach for NE-structure, we
find that the galaxies in the region of this overdensity are
significantly fainter and bluer than the galaxies in the cluster core
and in the E-structure, suggesting that the NE-structure is a background
cluster. However additional spectroscopic observations are necessary to
verify this point.

We have used the color-magnitude relation to analyze the galaxy contents in
structures S1, S2 and S3. We find that the galaxy populations in S1 and S2$+$S3
differ in its content. Most of the galaxies in S1 are redder than those
presented in S2 and S3, with an average color of \sloang$-$\sloanr$\sim$1.9. 
They lie well inside the red sequence for passive galaxies (see Fig. 7(a)). 
S2 and S3 are dominated by a population of blue, star-forming galaxies that,
in average, are $\sim0.3$mag fainter than the galaxies in S1. Further
evidence is provided by the fraction of emission-line galaxies in each
of the structure. We find that only 17\% of the galaxies in S1 are
emission-line galaxies, while the fraction of emission-line galaxies
in S2 and S3 is of the order of 31\%, in agreement with the galaxy
contents derived for both structures from the color-magnitude relation
analysis. 

We derived the X-ray temperature, the metal abundance and X-ray flux and 
luminosity in different energy bands using the plasma model described in
section 3.1. For the computed intra-cluster medium temperature of $kT
=3.3_{-0.6}^{+0.7}$~keV we find, using the $T_{X}-\sigma$ relation from
\citet{Xue00}, a velocity dispersion of $\sigma=672^{+57}_{53}$ km
s$^{-1}$. This value agrees well with the derived line-of-sight velocity
dispersion obtained for S1 (592$\pm$82 km s$^{-1}$), suggesting that the
X-ray emission is mainly associated to this structure. In addition, we find
that the cluster is slightly hotter than expected for its luminosity
($L^{bol}_{X}=11.8\pm0.9$ $10^{43}~ h_{70}^{-2}$ erg s$^{-1}$), compared to
the local $L_{X}$--$T_{X}$ correlation. However, its fall right on the
sub-sample relation at $z\approx0.34$ of \citet{Novicki02}.

We have used the weak-lensing analysis to map the mass distribution
in the area of the cluster. We find a good agreement between the
velocity dispersions derived from the \sloang and \sloanr using the
SIS model profile. However, there is a disagreement with the values
obtained from X-ray and from the kinematic of the member galaxies. We
used two fit models, the SIS and NFW, to compute the total mass of the
cluster. We find that the total mass inferred from weak-lensing of 2.1 to
3.7 $\times 10^{14} M_\odot$ at $r=0.5 h^{-1}_{70}$ Mpc (depending on the
band and the model adopted) is well in excess compared to the X-ray mass.
However, given the several difficulties for an accurate gravitational
lensing estimation in this field, particularly due to the
contribution of the other mass clumps and the uncertainty on the
average redshift of the background galaxies, these results should be taken
with caution. The presence of several sub-structures in  the X-ray, 
weak-lensing mass and galaxy density maps, the existence of a bridge of 
matter in the center of the cluster connecting different  sub-structures 
(detected in weak-lensing only) and the complex velocity distribution of 
member galaxies reveal that this cluster is dynamical active. Additional
evidence of the dynamical activity is given by the offsets we see
between the two brightest galaxies in the cluster core and the X-ray 
emission and the significantly offset between the velocity of the galaxy 
947 (the brigtest elliptical galaxy in S1) and the mean velocity of 
structure S1. Our main conclusion is that this poor cluster may be the 
core of a still forming rich cluster of galaxies. [VMF~98]~097 is in an 
environment with other nearby  substructures that, given their projected 
distance to the cluster, are probably gravitationally bound and will
eventually merge to form a rich cluster.

\acknowledgments

We would like to thank the anonymous referee for the useful comments 
and suggestions. ERC acknowledges the hospitality of the Departament 
of Astronomy of the Instituto de Astronomia, Geof\'{\i}sica e Ci\^encias
Atmosf\'ericas, Universidade de S\~ao Paulo, where this work was partially
done. ERC also acknowledges the support from CNPq through the PROSUL project.
GBLN, LSJ and CMO acknowledge support by the Brazilian agencies FAPESP and CNPq. 
We made use of the XMM-Newton arquival data: the XMM-Newton is an ESA
Science Mission with instruments and contribution direcly funded by ESA member
states and the USA (NASA) and the NASA/IPAC Extragalactic Database (NED) which
is operated by the Jet Propulsion Laboratory, California Institute  of
Technology, under contract with the National Aeronautics and Space
Administration.

\clearpage

\begin{deluxetable}{rcccccrrr}
\tabletypesize{\scriptsize}
\tablecaption{Galaxy radial velocities catalog\label{tab1}}
\tablenum{1}
\tablecolumns{9}
\tablewidth{0pc}
\tablehead{
\colhead{Galaxy id} &
\colhead{RA(2000)} &
\colhead{DEC(2000)} &
\colhead{r$^{'}_{0}$\tablenotemark{(c)}} &
\colhead{(g$^{'}-$r$^{'}_{0}$\tablenotemark{(c)})} &
\colhead{V$_{hel}$} &
\colhead{$\delta$v} &
\colhead{R} &
\colhead{\#lines}\\
\colhead{} &
\colhead{} &
\colhead{} &
\colhead{(mag)}&
\colhead{(mag)}&
\colhead{(km s$^{-1}$)}&
\colhead{(km s$^{-1}$)}&
\colhead{} &
\colhead{}}
\startdata
  514 & 11 17 18.76 & $+$07 42 42.3 & 20.49 & 1.98 & 183007 &   84 &  3.39 &\nodata  \\
 2492 & 11 17 19.55 & $+$07 46 01.0 & 22.63 & 1.28 & 143704 &   36 &  4.27 &\nodata  \\
 1937 & 11 17 19.36 & $+$07 45 07.7 & 19.41 & 0.94 & 109039 &   36 &\nodata  &  8   \\
  202 & 11 17 19.35 & $+$07 46 18.9 & 19.85 & 0.91 &  68022 &   10 &\nodata  &  9   \\
   21 & 11 17 19.76 & $+$07 47 03.2 & 19.11 & 1.46 & 107761 &   45 &  3.72 &\nodata \\
 1520 & 11 17 19.59 & $+$07 44 30.5 & 19.12 & 1.58 &  99787 &   40 &  9.06 &\nodata \\
 2079 & 11 17 21.80 & $+$07 45 18.4 & 22.56 & 0.98 & 144968 &   10 & \nodata & 8  \\
 2315 & 11 17 21.73 & $+$07 45 41.8 & 22.42 & 0.71 &  70683 &   59 & \nodata & 10  \\
  676 & 11 17 22.40 & $+$07 43 11.2 & 20.79 & 1.86 & 147672 &   61 &  4.60 &\nodata \\
  918 & 11 17 22.37 & $+$07 43 22.4 & 20.35 & 1.90 & 144136 &   61 &  5.12 &\nodata \\
 2691 & 11 17 22.81 & $+$07 46 23.4 & 21.25 & 1.46 & 170804 &   42 &  3.83 &\nodata \\
    1 & 11 17 22.69 & $+$07 45 31.4 & 17.11 & 1.09 &  38300 &   34 &\nodata&  6  \\
 2572 & 11 17 23.04 & $+$07 46 06.2 & 21.98 & 1.54 & 155596 &   35 &  4.81 &\nodata \\
 1570 & 11 17 23.02 & $+$07 44 35.3 & 21.19 & 1.32 & 144939 &   48 &  2.57 &\nodata \\
 1576 & 11 17 23.00 & $+$07 44 40.2 & 21.48 & 1.60 & 143740 &   46 &  3.70 &\nodata \\
 1568 & 11 17 23.29 & $+$07 44 31.9 & 21.36 & 1.68 & 143192 &   45 &\nodata&  8  \\
  135 & 11 17 23.44 & $+$07 46 36.5 & 20.73 & 1.75 & 202462 &   41 &  3.77 &\nodata \\
 2131 & 11 17 23.95 & $+$07 45 23.3 & 21.57 & 1.42 & 141945 &   52 &  4.75 &\nodata \\
 1299 & 11 17 24.29 & $+$07 44 10.6 & 19.65 & 1.72 & 147286 &   39 &  6.91 &\nodata \\
  190 & 11 17 24.14 & $+$07 46 40.8 & 20.94 & 1.15 & 103336 &   23 &     8 & \\
  694 & 11 17 24.48 & $+$07 42 58.2 & 21.53 & 1.94 & 144522 &   57 &  7.26 &\nodata \\
 3054 & 11 17 24.47 & $+$07 42 13.7 & 22.16 & 1.82 & 148235 &   61 &  4.57 &\nodata \\
  339 & 11 17 24.80 & $+$07 42 24.3 & 22.48 & 1.13 & 105922 &   13 &\nodata&  8  \\
  739 & 11 17 24.61 & $+$07 43 04.5 & 21.83 & 0.86 &  93069 &   20 &\nodata & 13  \\
  908 & 11 17 25.05 & $+$07 43 29.7 & 19.26 & 1.66 & 147157 &   48 &  4.23 &\nodata \\
 2973 & 11 17 25.22 & $+$07 42 04.6 & 20.46 & 1.94 & 145243 &   42 &  7.25 &\nodata \\
  441 & 11 17 25.54 & $+$07 42 34.0 & 22.37 & 0.91 & 146836 &   24 &\nodata&  8  \\
  910 & 11 17 25.92 & $+$07 43 34.5 & 19.80 & 0.92 &  90311 &   12 &\nodata& 15  \\
 1085 & 11 17 25.86 & $+$07 43 43.1 & 20.83 & 1.96 & 145337 &   56 &  5.35 &\nodata \\
  903 & 11 17 26.14 & $+$07 43 53.3 & 20.64 & 1.89 & 146283 &   39 &  8.36 &\nodata \\
 2156 & 11 17 26.14 & $+$07 45 29.7 & 20.59 & 1.88 & 145026 &   38 &  8.47 &\nodata \\
 1098 & 11 17 26.12 & $+$07 43 41.0 & 20.29 & 1.84 & 144645 &   49 &  5.36 &\nodata \\
  947 & 11 17 26.09 & $+$07 43 37.5 & 20.26 & 1.94 & 145882 &   47 &  6.11 &\nodata \\
 1205 & 11 17 26.48 & $+$07 43 53.1 & 22.10 & 1.23 & 149339 &   30 &  5.16 &\nodata \\
 1269 & 11 17 26.45 & $+$07 44 02.7 & 21.25 & 1.98 & 144528 &   24 &  7.79 &\nodata \\
 2346 & 11 17 26.28 & $+$07 45 48.3 & 20.99 & 1.32 & 103397 &   72 &\nodata& 10  \\
  904 & 11 17 26.75 & $+$07 43 24.0 & 18.90 & 0.94 &  48309 &   10 &\nodata& 10  \\
 1268 & 11 17 26.58 & $+$07 44 07.6 & 22.21 & 1.14 & 103535 &   26 &\nodata&  6  \\
  798 & 11 17 26.52 & $+$07 43 11.3 & 19.73 & 1.29 & 144724 &   44 &  7.50 &\nodata \\
  623 & 11 17 26.83 & $+$07 42 51.6 & 20.77 & 1.96 & 147448 &   57 &  6.08 &\nodata \\
 1939 & 11 17 26.79 & $+$07 45 08.1 & 20.88 & 1.77 & 170714 &   49 &  3.10 &\nodata \\
 1538 & 11 17 27.47 & $+$07 44 28.5 & 21.48 & 1.81 & 143797 &   51 &  5.85 &\nodata \\
 1153 & 11 17 27.46 & $+$07 43 50.1 & 21.18 & 1.86 & 183716 &   28 &  4.74 &\nodata \\
   10 & 11 17 27.45 & $+$07 46 57.2 & 17.57 & 1.30 &  47253 &   45 &  9.84 &\nodata \\
  121 & 11 17 27.42 & $+$07 46 47.7 & 19.15 & 1.27 &  47925 &   45 &  8.76 &\nodata \\
 1241 & 11 17 27.58 & $+$07 44 00.0 & 21.54 & 1.45 & 145443 &   21 &  7.72 &\nodata \\
  545 & 11 17 27.76 & $+$07 42 45.1 & 20.70 & 1.94 & 143211 &   50 &  9.12 &\nodata \\
  510 & 11 17 27.73 & $+$07 42 38.4 & 22.57 & 1.16 & 148356 &   76 &  3.94 &\nodata \\
    3 & 11 17 28.36 & $+$07 46 36.6 & 18.10 & 1.27 &  47375 &   29 & 11.94 &\nodata \\
   95 & 11 17 28.92 & $+$07 46 55.2 & 20.58 & 0.59 &  65862 &   15 &\nodata& 12  \\
 1611 & 11 17 29.06 & $+$07 44 37.3 & 21.46 & 1.12 & 149186 &   56 &\nodata&  8  \\
 1071 & 11 17 29.31 & $+$07 43 37.6 & 20.68 & 1.90 & 144779 &   18 & 12.24 &\nodata \\
 2422 & 11 17 29.22 & $+$07 45 52.1 & 21.09 & 1.89 & 149896 &   42 &  7.96 &\nodata \\
 2228 & 11 17 29.55 & $+$07 45 33.9 & 21.62 & 1.14 & 143249 &   64 &  4.05 &\nodata \\
    7 & 11 17 29.68 & $+$07 46 48.5 & 18.09 & 0.78 &  40722 &   10 &\nodata& 11  \\
    5 & 11 17 29.67 & $+$07 46 23.7 & 16.65 & 1.26 &  47854 &   43 &  8.03 &\nodata \\
    6 & 11 17 30.08 & $+$07 46 28.2 & 18.65 & 1.14 &  48516 &   26 &  8.75 &\nodata \\
  865 & 11 17 30.03 & $+$07 43 16.2 & 22.12 & 0.82 &  65836 &   25 &\nodata& 12  \\
   24 & 11 17 30.01 & $+$07 47 06.2 & 20.12 & 0.71 &  48649 &   18 &\nodata& 10  \\
  325 & 11 17 30.28 & $+$07 42 13.1 & 22.43 & 0.99 & 146910 &   26 &  4.18 &\nodata \\
 2105 & 11 17 30.19 & $+$07 45 23.3 & 20.27 & 1.28 & 149665 &   48 &  4.20 &\nodata \\
 1372 & 11 17 30.56 & $+$07 44 13.4 & 22.12 & 1.18 & 157677 &   19 &\nodata&  9  \\
  864 & 11 17 30.50 & $+$07 43 16.6 & 20.10 & 1.56 & 154346 &   40 &  6.53 &\nodata \\
 1099 & 11 17 30.42 & $+$07 44 04.2 & 20.51 & 1.89 & 144382 &   37 &  7.69 &\nodata \\
 2571 & 11 17 31.07 & $+$07 46 10.3 & 21.25 & 1.49 &  47531 &   56 &  3.29 &\nodata \\
  988 & 11 17 30.99 & $+$07 43 29.6 & 19.70 & 1.04 & 107733 &   64 &\nodata&  8  \\
 1793 & 11 17 30.86 & $+$07 44 52.4 & 21.89 & 0.56 & 300960 &  114 &\nodata&  7  \\
 2921 & 11 17 31.82 & $+$07 41 57.9 & 21.21 & 2.01 & 153491 &   50 &  5.12 &\nodata \\
 1811 & 11 17 32.28 & $+$07 44 55.3 & 21.14 & 1.80 & 157625 &   32 &  6.60 &\nodata \\
 1219 & 11 17 32.42 & $+$07 43 58.0 & 21.66 & 1.94 & 145199 &   44 &  6.52 &\nodata \\
  663 & 11 17 33.43 & $+$07 43 01.3 & 18.79 & 0.41 &  12677 &   16 &\nodata& 13  \\
  525 & 11 17 34.31 & $+$07 42 40.0 & 22.30 & 1.92 & 184040 &   57 &  4.91 &\nodata \\
  359 & 11 17 35.53 & $+$07 42 29.2 & 21.30 & 1.47 & 147726 &   60 &  3.62 &\nodata \\
  313 & 11 17 35.73 & $+$07 42 52.6 & 19.99 & 1.98 & 143826 &   35 &  5.80 &\nodata \\
 1719 & 11 17 35.89 & $+$07 44 59.7 & 19.33 & 0.68 &  33815 &   41 &\nodata& 12  \\
\enddata
\tablecomments{The units of Right Ascension are hours, minutes and
seconds, and those of Declination are degrees, arcminutes and arcseconds.}
\tablenotetext{(a)}{Total magnitudes and colors are corrected by galactic
extinction using the maps of \citet[see Sect.~\ref{sec:colmag}]{sch98}}
\end{deluxetable}

\clearpage

\begin{deluxetable}{ccccc}
\tabletypesize{\scriptsize}
\tablenum{2}
\tablecaption{Summary of XMM observations. \label{tbl:resumoXMM}}
\tablewidth{0pt}
\tablehead{
\colhead{obs. date}  & \colhead{obsID}     & \colhead{detector} &
\colhead{net exp. time} & \colhead{net}  \\
\colhead{}  & \colhead{} & \colhead{} & \colhead{(ks)}&
\colhead{counts} }
\startdata
    2001-11-25 & 082340101 & MOS1     & 47.5  &  252 \\
    2001-11-25 & 082340101 & PN       & 40.3  &  611 \\
    2004-06-10 & 203560201 & MOS1     & 59.2  &  322 \\
    2004-06-26 & 203560401 & MOS1     & 63.8  &  389
\enddata
\end{deluxetable}

\vspace{1cm}

\begin{deluxetable}{ccccc}
\tabletypesize{\scriptsize}
\tablecaption{X-ray flux and luminosities in different energy bands (in keV).\label{tbl:LX}}
\tablenum{3}
\tablewidth{0pc}
\tablehead{
\multicolumn{2}{c}{Flux} & \multicolumn{3}{c}{Luminosity}\\
\colhead{[0.5$-$2.0]} & \colhead{[2.0$-$10.0]} & \colhead{[0.5$-$2.0]} & \colhead{[2.0$-$10.0]} & \colhead{bolom.}}
\startdata
$5.25\pm0.43$ & $3.61\pm0.30$ & $4.19\pm0.35$ & $4.44\pm0.37$ & $11.8\pm0.9$ \\
\enddata
\tablecomments{Flux is in units of $10^{-14}$ erg s$^{-1}$ cm$^{-1}$.
Luminosity, measured
in the source rest frame, is in units of $10^{43}~ h_{70}^{-2}$ erg s$^{-1}$.}
\end{deluxetable}

\vspace{1cm}

\begin{deluxetable}{rccccccc}
\tabletypesize{\scriptsize}
\tablecaption{Dynamical parameters of the [VMF~98]~097 structures \label{tab:vel}}
\tablenum{4}
\tablecolumns{8}
\tablewidth{0pc}
\tablehead{
\colhead{Structure} &
\colhead{RA(2000)} &
\colhead{DEC(2000)} &
\colhead{$N_{\rm mem}$} &
\colhead{C$_{\rm BI}$} &
\colhead{S$_{\rm BI}$} &
\colhead{$R_{\rm vir}$} &
\colhead{$M_{\rm vir}$}\\
\colhead{} &
\colhead{} &
\colhead{} &
\colhead{} &
\colhead{(km s$^{-1}$)} &
\colhead{(km s$^{-1}$)} &
\colhead{($h_{70}^{-1}$~Mpc)} &
\colhead{($10^{14}$ M$_\odot$ h$_{70}^{-1}$)}}
\startdata
S1   & 11 17 26.4 & $+$07 44 01.3 & 23 & 0.48218$\pm$0.00042 & 592$\pm$82  & 1.02$^{+0.13}_{-0.12}$ & 1.05$^{+0.23}_{-0.10}$ \\[6pt]
S2   & 11 17 26.8 & $+$07 42 52.4 &  9 & 0.49191$\pm$0.00048 & 391$\pm$85  & \nodata                & \nodata                \\
\enddata
\end{deluxetable}

\vspace{1cm}

\begin{deluxetable}{rcccccc}
\tabletypesize{\scriptsize}
\tablecaption{Median magnitudes and colors of the main overdensity
regions in [VMF~98]~097 cluster \label{tab:mags}}
\tablenum{5}
\tablecolumns{3}
\tablewidth{0pc}
\tablehead{
\colhead{Overdensity} &
\colhead{\# of galaxies} &
\colhead{$\langle$ \sloanr $\rangle$ ($R \le 20$\arcsec)} &
\colhead{$\langle$ (\sloang $-$ \sloanr) $\rangle$ ($R \le 20$\arcsec)} \\
\colhead{} &
\colhead{} &
\colhead{(mag)} &
\colhead{(mag)}}
\startdata
cluster core & 14 & 20.67 & 1.64 \\
E-structure  & 11 & 21.81 & 1.51 \\
NE-structure &  7 & 22.25 & 1.11 \\
\enddata
\end{deluxetable}

\vspace{1cm}

\begin{deluxetable}{lclc}
\tabletypesize{\scriptsize}
\tablewidth{0pt}
\tablenum{6}
\tablecaption{Model fitting of the weak-lensing data. \label{weak_results}}
\tablehead{
\colhead{(1)} & \colhead{(2)} & \colhead{(3)} & \colhead{(4)}\\
\colhead{Method} & \colhead{Filter}  &
\colhead{Fitted Parameters} & \colhead{M($r<0.5$ Mpc)}\\
\colhead{}  & \colhead{} & \colhead{} &
\colhead{($10^{14}$ M$_\odot$ h$_{70}^{-1}$)}}
\startdata
SIS        & \sloang ~  & $\sigma=809\pm89$ \kms                & $2.4\pm0.5$\\
SIS        & \sloanr ~  & $\sigma=746\pm86$ \kms                & $2.1\pm0.4$\\
NFW($c=5$) & \sloang ~  & $R_{200}=2.6\pm0.4$ Mpc               & $3.7\pm0.4$\\
NFW($c=5$) & \sloanr ~  & $R_{200}=2.4\pm0.3$ Mpc               & $3.4\pm0.4$\\
\enddata
\end{deluxetable}

\end{document}